\documentclass[sigconf, nonacm]{acmart}

\AtBeginDocument{%
  }

\acmConference[Conference acronym 'XX]{conference title}{
  2026}{}
\acmISBN{}

\usepackage{longtable}
\usepackage{caption}
\usepackage{float}
\usepackage{multirow}  

\usepackage{amsmath}
\usepackage{algorithm}
\usepackage{algorithmic}
\usepackage{float}
\newtheorem{theorem}{Theorem}
\newtheorem{corollary}[theorem]{Corollary}

\begin{document}

\title{Beyond Basic A/B Testing: Improving Statistical Efficiency for Business Growth}

\author{Changshuai Wei}
\authornote{Corresponding Author. Theoretical development was primarily carried out by C. Wei, who also led the
design, execution, and writing of the manuscript. All authors contributed to algorithm implementation, simulation studies, real-world applications, and manuscript drafting.}
\email{chawei@linkedin.com}
\affiliation{%
  \institution{LinkedIn Corporation}
  \city{Seattle} 
  \country{USA}
}

\author{Phuc Nguyen}
\email{honnguyen@linkedin.com}
\affiliation{%
  \institution{LinkedIn Corporation}
  \city{Sunnyvale} 
  \country{USA}
}

\author{Benjamin Zelditch}
\email{bzelditch@linkedin.com}
\affiliation{%
  \institution{LinkedIn Corporation}
  \city{Sunnyvale}
  \country{USA}
}

\author{Joyce Chen}
\email{joychen@linkedin.com}
\affiliation{%
  \institution{LinkedIn Corporation}
  \city{Sunnyvale}
  \country{USA}
}

\renewcommand{\shortauthors}{Wei et al.}

\begin{abstract}
The standard A/B testing approaches are mostly based on t-test in large scale industry applications. These standard approaches however suffers from low statistical power in business settings, due to nature of small sample-size or non-Gaussian distribution or return-on-investment (ROI) consideration. In this paper, we (i) show the statistical efficiency of using estimating equation and U statistics, which can address these issues separately; and (ii) propose a novel doubly robust generalized U that allows flexible definition of treatment effect, and can handles small samples, distribution robustness, ROI and confounding consideration in one framework. We provide theoretical results on asymptotics and efficiency bounds, together with insights on the efficiency gain from theoretical analysis. We further conduct comprehensive simulation studies, apply the methods to multiple real A/B tests at LinkedIn, and share results and learnings that are broadly useful.
\end{abstract}




\maketitle
\newcommand\kddavailabilityurl{https://doi.org/10.5281/zenodo.18064926}
\ifdefempty{\kddavailabilityurl}{}{
\begingroup\small\noindent\raggedright\textbf{Resource Availability:}\\
The source code of this paper has been made publicly available at \url{\kddavailabilityurl} (i.e., \href{https://github.com/linkedin/robustInfer/tree/v0.1.0}{\textit{robustInfer}} v0.1.0 release).
\endgroup
}
\sloppy

\section{Introduction}
\label{sec:intro}


Controlled experiments have been the gold standard of measuring the effect of a treatment/drug in biological and medical research for more than 100 years \cite{fisher1928statistical, fisher1935design}. In the last few decades, the rise of the internet and machine learning (ML) algorithms led to the development and revival of controlled experiments for online internet applications, i.e., A/B testing\cite{kohavi2009}. Most of the A/B testing in industry follows standard statistical approaches, e.g., t-test, particularly in large-scale recommender systems (e.g., Feeds, Ads, Growth), which involve sample sizes on the order of millions to billions, and measure engagement metrics such as clicks or impressions. 

In business settings, e.g., Marketing, Software-as-a-Service (SaaS), and Business-to-Bussiness
(B2B), there are unique challenges, where standard approaches like the t-test can lead to either incorrect conclusions or insufficient statistical power: (i) \textit{Return-on-Investment} (ROI) or Return-on-Ad-Spend (ROAS) type of measurement is almost always key consideration in business setting. There has been little research on how to efficiently measure this type of metric in the A/B testing setting; (ii) \textit{Small sample sizes} are very common in business‐setting A/B tests, since increasing the sample size typically incurs additional cost; (iii) Revenue, as a core metric in business setting, is typically right-skewed with a \textit{heavy tail}. Since revenue generation is typically sparse event conditioning on sales outreach or marketing touch-point, we also need to address \textit{zero-inflation}.

In this paper, we apply a series of statistical methods to address the above challenges to improve statistical efficiency. We further develop a novel doubly robust generalized U statistic that combines advantage of existing methods. As far as we know, this is among the first attempts to unify efficient statistical methods for A/B test, particularly for business setting. The key contributions of the paper are:\\
1) Application innovations on using Estimating Equations and U statistics to improve statistical efficiency in AB test. \\
2) \textbf{Theoretical development} on (i) systematic analysis of \textit{asymptotic efficiency} for Estimating Equation and U statistics, and more importantly (ii) a novel \textit{doubly robust \textbf{generalized U}} that attains the \textit{semi-parametric efficiency bound} and can concurrently address ROI, longitudinal analysis, and ill-behaved distributions, as well as (iii) \textit{rigorous efficient algorithms} for large data for broader applicability. \\
3) We conducted \textit{thorough simulation studies} to evaluate the empirical efficiencies and applied the methods to \textit{multiple real business applications}.

In-depth discussion on application innovations and theoretical  contributions can be found in section~\ref{sec:Discn}. Though these methods are proposed to address challenges in business setting, they are broadly applicable to general A/B test. 

The rest of the paper is structured as follows. For the remainder of section~\ref{sec:intro}, we'll discuss related work and introduce the problem setup and preliminaries. In section~\ref{sec:ee}, we discuss efficiency of estimating equations. In section~\ref{sec:NonGausDist}, we discuss efficiency of U statistics and introduce Zero-Trimmed U. In section~\ref{sec:AdvDistFree}, we develop methodology for doubly robust generalized U test. Then, we conduct simulation studies and real data analysis in section~\ref{sec:SimuApp} and conclude the paper with discussion in section ~\ref{sec:Discn} and section~\ref{sec:conclusion}. Details on algorithms, theoretical proof, analytical derivation, and simulation set-up can be found in \hyperref[sec:apx]{Appendix}.

\subsection{Related Works}

There have been multiple research efforts to address limitations of standard t-tests in A/B test, particularly for low sensitivity and small treatment effects \cite{larsen2024statistical}. Covariate adjustment\cite{fisher1928statistical} has been widely used as an improvement to t-test or proportion test in biomedical experiments\cite{ hernandez2004,freedman2008regression,kahan2014risks}. An important relevant development in A/B tests is Controlled-experiment Using Pre-Experiment Data (CUPED)\cite{deng2013}, which leverages pre-experiment metrics in a simple linear adjustment to reduce variance. Later extension of the methods includes leveraging in-experiment data, non-linear predictive modeling, and individual-variance weighting for further reduction of variance\cite{poyarkov2016, liou2020, deng2023}. Meanwhile, there are increasing concerns on other challenges, such as repeated measurements\cite{liou2022privacy, zhou2023all} and non-Gaussian heavy-tailed distributions \cite{Jiang2020, azevedo2020ab}. Semi-parametric approaches such as GEE have been well adopted in biomedical field for repeated measurements \cite{liang1986longitudinal, wang2014generalized}. Nonparametric methods, such as Wilcoxon Rank-sum and U-statistic, can provide robustness to ill-behaved distribution\cite{mann1947test, hoeffding1948u, kowalski2008}. In recent years, U statistics have emerged as an important class of statistical methods in biomedical research \cite{ma2012inference, mao2018causal,yin2024highly} and social sciences\cite{chen2014,ai2020mann,mao2024}, with particular developments in genomics \cite{wei2017generalized,wei2020multi,wang2023} and causal inferences\cite{wu2014,vermeulen2015increasing,chen2024} for public health studies. The application of U statistics in tech industry are largely limited to ROC-AUC (equivalent to Mann-Whitney U \cite{hanley1982meaning}) for ML models' evaluation, and it's often just used as point estimate. While there are some development on metric learning and non-directional type of tests(e.g., goodness of fit, independence) using U statistics\cite{clemenccon2016scaling, schrab2022efficient, schrab2023mmd}, they are not suitable for A/B testing. 

\subsection{Problem Setup and Preliminaries}

Let's assume we perform A/B test to compare two treatment $z=0$ vs $z=1$ on primary metric $y$. Meanwhile, we also observe vector of other metrics $w$ (e.g., cost).
Our goal is to evaluate "improvement" of $y$ from the treatment over control group (directional test).

\textbf{\textit{T Test}}: One common formulation of the "improvement" is: $\delta = E(y_{i1} - y_{i0})$, and we can use t-test for the corresponding null vs alternative hypotheses:
$H_{0}: \delta=0, \text{ vs  } H_1: \delta>0$. The corresponding t-statistics is $t_n = \frac{\bar{y}_1 - \bar{y}_0}{\sqrt{\hat{v}_{10}}}$, where, $\bar{y}_k$ is sample mean for $z_i=k$, and $\hat{v}_{10}$ is corresponding variance estimator depending on equal or unequal variance assumption. 

\textbf{\textit{Statistical Efficiency}}: We can measure the statistical efficiency of a estimation process by mean squared error (MSE), and define the relative efficiency by inverse ratio of MSE,
$r_n(\hat{\delta}_1,\hat{\delta}_2) = \frac{E(\hat{\delta}_2 - \delta)^2}{E(\hat{\delta}_1 - \delta)^2}=\frac{Var(\hat{\delta}_2) + Bias^2(\hat{\delta}_2)}{Var(\hat{\delta}_1) + Bias^2(\hat{\delta}_1)}$,
where $\hat{\delta}_1$ and $\hat{\delta}_2$ are two different estimator of $\delta$. When both estimator are unbiased, the relative efficiency reduced to ratio of variance. We can define asymptotic relative efficiency (ARE) as $r(\hat{\delta}_1,\hat{\delta}_2) = \lim_{n\rightarrow \infty} r_n(\hat{\delta}_1,\hat{\delta}_2)$. 

For hypotheses testing, we can use Pitman efficiency, 
$r(t_1, t_2) = \lim_{n\rightarrow \infty} \frac{n_{t_2}}{n_{t_1}}$,
where $n_{t_1}$ and $n_{t_2}$ are sample size required to reach the same power $\beta$ for $\alpha$ level test, with test statistic $t_1$ and $t_2$ respectively. Assume local alternative (e.g., small location shift $\delta$), and asymptotic normality of test statistics (i.e., $\sqrt{n}t_{n,i}\rightarrow_d N(\mu_i(\delta), \sigma^2(\delta))$), Pitman efficiency is equivalent to the following alternative definition of efficiency:
$r(t_1, t_2) = \frac{\lambda_1^2}{\lambda_2^2} = \left(\frac{\mu_1'(0)/\sigma_1(0)}{\mu_2'(0)/\sigma_2(0)}\right)^2$, 
where $\lambda_k = \frac{\mu_k'(0)}{\sigma_k(0)}$ is slope of test $k$. The equivalence can be shown observing the power function $\beta(\delta) = 1 - \Phi(z_{\alpha} - \sqrt{n}\delta\lambda)$, and thus $n\propto \frac{1}{\lambda}$.

In this paper, we will evaluate statistical efficiency of a series of existing and new methodologies. The comparison and evaluation will be either asymptotic efficiency in analytic form or empirical efficiency in terms of simulation studies. 

\section{Estimating Equation}
\label{sec:ee}
\subsection{Efficiency with Covariate Adjustment}
\label{sec:reg}

Cost is core guardrail in evaluation of algorithm or strategies in business setting. One common strategy is to perform t-tests on both primary metrics (e.g., revenue) and guardrail metrics (e.g., cost), and make decisions on heuristic combination of the two t-test. However, this type of strategy lacks a unified view on ROI and can lead to decision confusion when the conclusion on the two metric goes opposite way. 

Regression adjustment approach\cite{fisher1928statistical, freedman2008regression} can be used as a fundamental approach for measuring ROI. We can form a regression model: $y_i = \beta_0+\beta_1z_i+\gamma^T w_i +\epsilon_i$, and $\beta_1$ can represent "ROI", i.e., impact on revenue adjusted for cost. Further, regression adjustment provide advantages on statistical efficiency compared with t-test\cite{freedman2008regression, lin2013agnostic}.

Under confounding and above parametric set-up, $\beta_1$ is unbiased, while t-test is biased by a constant term $\gamma^T\left [E(w|z=1)-E(w|z=0)\right ]$. In this case, the ARE is dominated by the bias term, and hence $r(\hat{\beta}_1, \hat{\tau})\rightarrow \infty$ as $n\rightarrow \infty$. 

When there are no confounding, i.e., $z \perp w$, both are unbiased with $r(\hat{\beta}_1, \hat{\tau}) = 1 + \frac{\sigma^2_{w}}{\sigma^2} \geq 1$, where $\sigma^2_w = \gamma^TVar(w)\gamma$. As long as $w$ can explain some variance of $y$ (i.e., $\sigma^2_{w}>0$ or $\gamma \neq 0$), regression adjustment is strictly more efficient than t-test. This is also the key reason behind efficiency of all the CUPED type of methods, basically by including pre-experiment variables $w$ that can explain some variance of $y$ and satisfy $z \perp w$ by design.

\subsection{Efficiency with Repeated Measurement}
\label{sec:LA}
For almost all A/B testing in industry, metrics are monitored regularly over time. This is a unique characteristic in A/B test: repeated measurement of metrics have negligible (additional) cost, whereas in other fields like biomedical field, repeated measurements are often constrained by expense. 

Therefore, it is essential to leverage the repeated measurement in A/B testing to improve efficiency\cite{diggle2002analysis}. Instead of common practice where analysis is performed on a snapshot of data, GEE\cite{liang1986longitudinal, wang2014generalized} can be used to analyze the full longitudinal data. We can form a repeated measurement regression model, $y_{it} = \beta_0+\beta_1z_i+\gamma^T w_{it} +\epsilon_{it}$, where $t$ represent the time index, $ Cov(\epsilon_i)=\sigma^2R$, and $R \succ 0$. 

Since GEE uses all the data, intuitively it'll have higher efficiency. For ease of comparison, we assume $w_{it}$ is constant overtime, i.e., $w_{it}=w_i$. 
We know variance of GEE estimate, 
$Var(\hat{\theta}_{gee})= \frac{\sigma^2}{e^TR^{-1}e}(\sum_i x_i x_i^T)^{-1}$,
where, $x_i = [1, z_i, w_i^T]^T$, $e=[1,1,\cdots,1]^T$, and $X_i = e x_i^T$.
For the snapshot regression analysis, we know,
$Var(\hat{\theta}_{reg}) = \sigma^2(\sum_i x_i x_i^T)^{-1}$.
Then the relative efficiency is $r(\hat{\beta}_{1,gee}, \hat{\beta}_{1,reg}) = e^TR^{-1}e $.

When $R \succ 0$, i.e., positive definite, we have $eR^{-1}e^T > 1$, which means that GEE is strictly more efficient than regression adjustment. When $R = ee^T$, i.e., perfect correlation among repeated measurement, $r(\hat{\beta}_{1,gee}, \hat{\beta}_{1,reg})=1$. In fact, let $\bar{\rho}$ be the average correlation among repeated measurement, i.e., $\bar{\rho} = \frac{1}{T(T-1)}\sum_{i\neq j}R_{ij}$, we know
$r(\hat{\theta}_{gee}, \hat{\theta}_{reg})\geq \frac{T}{1+(T-1)\bar{\rho}}$. (Appendix~\ref{apx:GEE_efficiency})


\section{U Statistics}
\label{sec:NonGausDist}

In many common business scenarios, primary metrics such as revenue exhibits strong characteristics of Non-Gaussian distributions, e.g., right skewed heavy tailed distribution. Further, important business event such as conversions happens sparsely, making the primary metrics often zero inflated. In these scenarios, standard parametric approach such as $t$-test can suffers from inflated type I error or power loss. 

\subsection{Efficiency of Mann-Whitney U}
\label{sec:mwU}

Given two independent samples $\{y_{1i}\}_{i=1}^{n_1}$ and $\{y_{0j}\}_{j=1}^{n_0}$, the Mann-Whitney U statistic (MWU) \cite{mann1947test, hoeffding1948u} is given by $U = \frac{1}{n_0n_1} \sum_{i=1}^{n_1} \sum_{j=1}^{n_0} I_{y_{1i} \geq y_{0j} }$, where $I$ is indicator function. $U$ is an unbiased estimator for $\delta = P(y_{1i}>y_{0j})$ and we can perform a score type of test with $\sigma_u^2 = \frac{n_0+n_1}{12}(\frac{1}{n_0}+\frac{1}{n_1})$.


Let $\kappa(y_{1i})$ denote the rank of $y_{1i}$ in the combined sample of $\{y_{1i}\}_{i=1}^{n_1}$ and $\{y_{0j}\}_{j=1}^{n_0}$ in descending order, i.e., $\kappa(y_{1i}) = 1+\sum_{i'\neq i}^{n_1} I_{y_{1i} < y_{1i'}} + \sum_j^{n_0} I_{y_{1i}<y_{0j}}$. The Wilcoxon rank-sum test statistic is given by $W = \sum_{i=1}^{n_1}\kappa(y_{1i}) - \frac{n_1(n_1+n_0+1)}{2} = -n_1n_0U + \frac{n_1n_0}{2}$. This relationship between $W$ and $U$ allows us to compute $U$ efficiently for large sample sizes by leveraging fast ranking algorithms. 

The Pitman relative efficiency of MWU vs t-test \cite{blair1980, bridge1999} is: $r(U,\tau) = 12\sigma^2\left[\int f^2(x)dx\right ]^2$, where $f$ represent the underlying distribution.

Using this result, we know for normal distribution, $r(U,\tau)=\frac{3}{\pi}$; for Laplace distribution, $r(U,\tau)=1.5$; for log-normal $r(U,\tau) = \frac{3}{\pi b^2}(e^{\frac{5}{2}b^2}-e^{\frac{3}{2}b^2})$, which increase exponentially with variance parameter of log-normal; and for Cauchy distribution $r(U,\tau)=\infty$ as t-test will break. (Appendix~\ref{sec:UPitmanExample})

\subsection{Zero-Trimmed U Test}
\label{sec:TrimU}
The challenges of non-Gaussian distribution is often two fold in business scenario, the heavy tail nature and the zero-inflation nature. We can exploit the zero-inflation characteristic to further improve efficiency. The idea is to trim off the same proportions of excessive zero and focus on the the continuous distributed part and “residual” zero difference.



Let $n_0^+=\sum_{i=1}^{n_1}I_{y_{1i}>0}$, and $n_1^+=\sum_{j=1}^{n_0}I_{y_{0j}>0}$. We can get proportion of positive values in the two samples: $\hat{p}_{1} = \frac{n_1^+}{n_1}$  and $\hat{p}_{0} = \frac{n_0^+}{n_0}$, and define $\hat{p} = \max\{\hat{p}_{1}, \hat{p}_{0}\}$. Remove $n_1(1-\hat{p})$ zeros from $\{y_{1i}\}_{i=1}^{n_1}$ and $n_0(1-\hat{p})$ zeros from $\{y_{0j}\}_{j=1}^{n_0}$. Let $\{y'_{1i}\}_{i=1}^{n'_1}$ and $\{y'_{0j}\}_{j=1}^{n'_0}$ denote the residual samples containing $n'_1=n_1\hat{p}$ and $n'_0=n_0 \hat{p}$ data points, respectively. Let $\kappa(y'_{1i})$ denote the rank of $y'_{1i}$ in the combined residual samples in descending order.   

The zero-trimmed U (ZTU) is given by $W' = \sum_{i=1}^{n'_1}\kappa(y'_{1i}) - \frac{n'_1(n'_1+n'_0+1)}{2}$. Conditioning on $\hat{p}_0$ and $\hat{p}_1$, we have $\mathbb{E}(W'|\hat{p}_{1}, \hat{p}_{0}) = \frac{n'_1n_0^+ - n_1^+n'_0}{2}$ and $\textrm{Var}(W'|\hat{p}_{1}, \hat{p}_{0}) = \frac{n^+_0n^+_1(n^+_0+n^+_1+1)}{12}$. Then we can show (details in Appendix~\ref{sec:ZtuPitAsym}) its variance as: $\sigma^2_{W'} = \frac{n_1^2n_0^2}{4}\hat{p}^2(\frac{\hat{p}_1(1-\hat{p}_1)}{n_1}+\frac{\hat{p}_0(1-\hat{p}_0)}{n_0}) + \frac{n_1n_0\hat{p}_1\hat{p}_0}{12}(n_1\hat{p}_1+n_0\hat{p}_0) + o_p(n^3)$. We can estimate the variance empirically,
$\hat{\sigma}^2_{W'}=\frac{n_1^2n_0^2}{4}\hat{p}^2(\frac{\hat{p}_1(1-\hat{p}_1)}{n_1}+\frac{\hat{p}_0(1-\hat{p}_0)}{n_0}) + \frac{n^+_0n^+_1(n^+_0+n^+_1)}{12}$
and perform statistical testing via $\frac{W'}{\hat{\sigma}_{W'}}$. 

Alternatively, we can use MWU without trimming off zeros. Let $W^o$ denote standard MWU, and $W$ the tie-adjusted MWU.

To facilitate comparison of efficiency, we can assume $m=p_1-p_0$ and $d=P(y_1^+>y_0^+)-\frac{1}{2}$. The compound hypothesis would be: $H_0: m=0, \text{ and }d=0$; $H_1: (1-I_{m>0})(1-I_{d>0}) =0$. We state the following theorem for Pitman efficiency of ZTU vs MWU.

\begin{theorem}\label{thm:Zero_Eff}
Let p denote proportion of positive values under $H_0$, $\phi$ be the direction of compound $H_1$, $\nu$ be the effect size along direction $\phi$, i.e., $m(\nu) = \nu \cos\phi$ and $d(\nu)=\nu \sin\phi$, The compound hypothesis can be transformed to simple hypothesis testing with direction of $\phi$, i.e., $H_0: \nu=0 \text{, vs }H_1^{\phi}:\nu>0$. 
And the corresponding Pitman efficiencies are,
\begin{align}
    r^{\phi}(W',W^o) 
    =\frac{\frac{1}{3}}{p^3-p^4+\frac{p^3}{3}}\left ( \frac{p\cos \phi+2p^2\sin \phi}{\cos \phi + 2p^2 \sin \phi} \right )^2, \label{eq:r_ww_o}
\end{align}
\begin{align}
    r^{\phi}(W',W) 
    =\frac{1-p+\frac{p^2}{3}}{p^2-p^3+\frac{p^2}{3}}\left ( \frac{p\cos \phi+2p^2\sin \phi}{\cos \phi + 2p^2 \sin \phi} \right )^2.
\end{align}
\end{theorem}

We provide proof in Appendix~\ref{sec:ZtuPitEff}. With above theorem, we can investigate the relative efficiency by varying value of $p \in (0,1]$ and $\phi \in [0,\frac{\pi}{2}]$. 

\begin{figure}[h]
  \centering
  \includegraphics[width=0.45\textwidth]{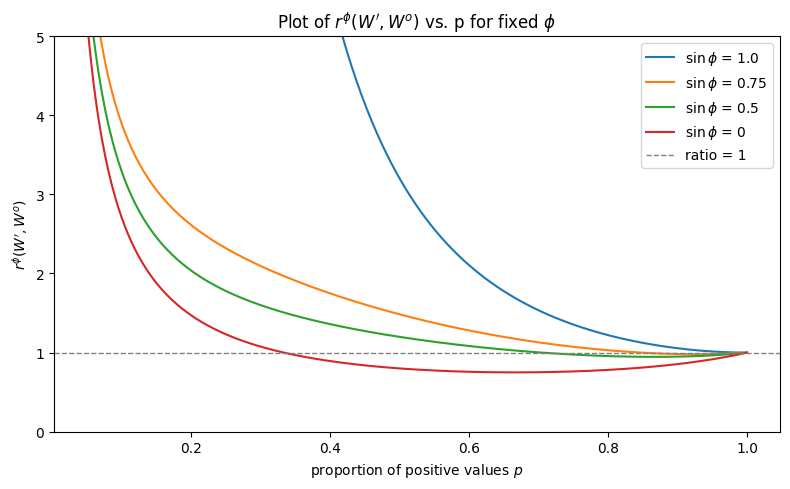}
  \includegraphics[width=0.45\textwidth]{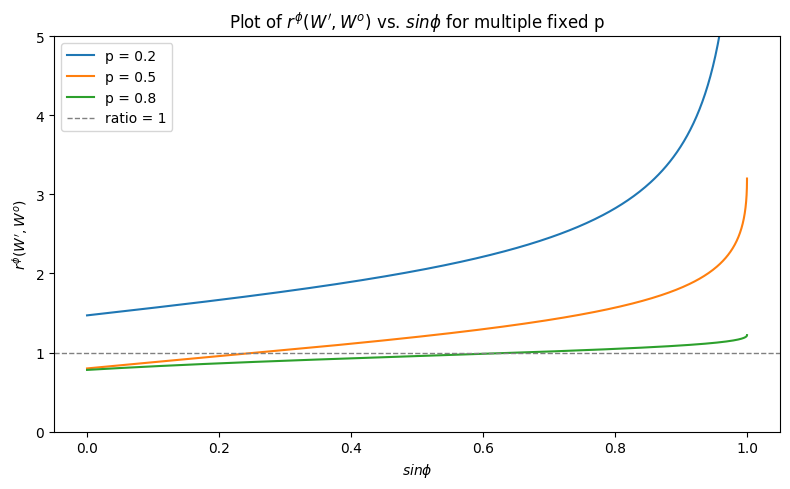}
  \caption{Plot of \(r^{\phi}(W',W^o)\) }
  \label{fig:rphi-o}
\end{figure}

\begin{figure}[h]
  \centering
  \includegraphics[width=0.45\textwidth]{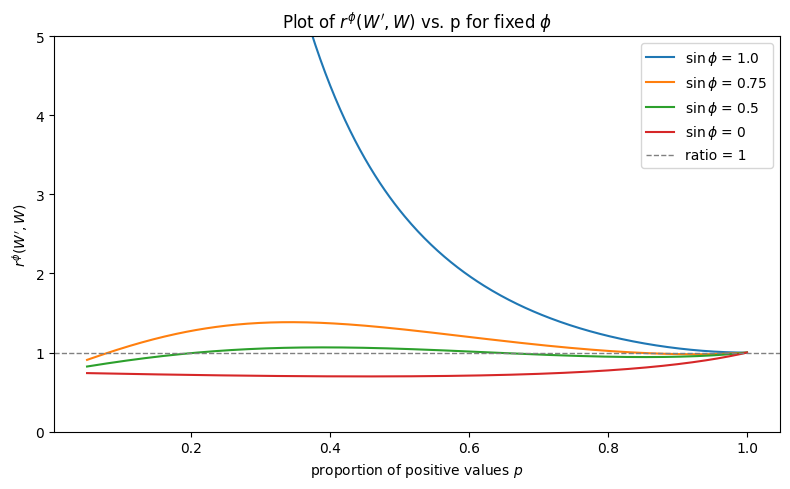}
  \includegraphics[width=0.45\textwidth]{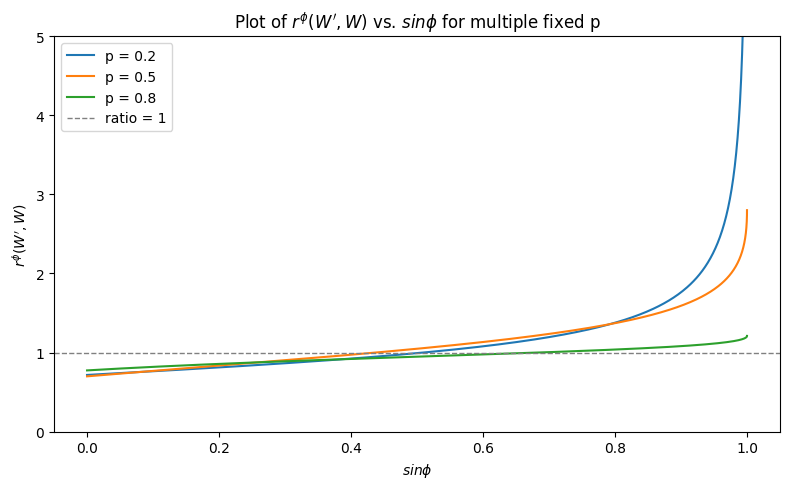}
  \caption{Plot of \(r^{\phi}(W',W)\).}
  \label{fig:rphi}
\end{figure}

From Figure~\ref{fig:rphi-o}, we can see ZTU is more efficient that MWU for most of situations. For Figure~\ref{fig:rphi}, we can see ZTU is comparable to tie-adjusted-MWU and significantly more efficient in many situations.

\section{U statistics with Estimating Equation}
\label{sec:AdvDistFree}
We develop general and robust U statistics based methodology in this section that can (i) measure various definitions of treatment effect, (ii) address both covariate adjustment and "ill-behaved" distribution in business setting, and (iii) can also utilize repeated measurements in A/B tests.

\subsection{Doubly Robust Generalized U Test}
\label{sec:DR_U}

Let $y_i$ denote the response variable that measure the business return, e.g., conversion or revenue, $z_i$ denote the treatment assignment, and $w_i$ denote the variables that needs to be adjusted, e.g., cost or impression. We define the treatment effect as, $\delta = E(\varphi(y_{i1}-y_{i0}))$,
where, $y_{i1}$ and $y_{i0}$ represent response variables for $z_i=1$ and $z_i=0$ respectively. Obviously, we observe only one of $y_{i0}$ and $y_{i1}$. 

$\varphi(\cdot)$ is a monotonic function with finite second moment, i.e., $E(\varphi^2(y_{i1}-y_{i0}))<\infty$. For example, when $\varphi(y_{i1}-y_{i0})=I_{y_{i1}>y_{j0}}$, we know $\delta = P(y_{i1}>y_{i0})$. We can also use other monotonic finite function like logistic function, $\varphi(y_{i1}-y_{i0}) = [1+\exp(-(y_{i1}-y_{i0}))]^{-1}$, Probit function, $\varphi(y_{i1}-y_{i0}) = \Phi(y_{i1}-y_{i0})$ or signed Laplacian kernel, $\varphi(y_{i1}-y_{i0}) = sign(y_{i1}-y_{i0})\exp(-\frac{y_{i1}-y_{i0}}{\sigma})$. Note when $\varphi(\cdot)$ is identity, we get $\delta = E(y_{i1}-y_{i0})$, which is treatment effect corresponding to t-test. However, it doesn't guarantee finite second moment condition(e.g., infinite second moment under Cauchy distribution).

Let $p=E(z)$. We can define a generalized U statistics:
$U_n = \left[ \binom{n}{2}\right]^{-1} \sum_{i,j\in C^n_2 } h(y_i,y_j)$,
where,
$h(y_i, y_j) = \varphi(y_{i1}-y_{j0})\xi_{ij} + \varphi(y_{j1}-y_{i0})\xi_{ji}$, and $\xi_{ij}=\frac{z_i(1-z_j)}{2p(1-p)}$. When there are no confounding, we know $E(U_n)=\delta$. In fact, when $\varphi(y_{i1}-y_{i0})=I_{y_{i1}>y_{j0}}$, it is equivalent to MWU.

To address covariate adjustment, let $\pi_i = E(z_i|w_i)$, and $g_{ij}=E(\varphi(y_{i1} - y_{j0})|w_i, w_j)$. We can form a efficient \textbf{d}oubly \textbf{r}obust\cite{robins1994estimation, hirano2003efficient, tsiatis2006semiparametric} version of the \textbf{g}eneralized \textbf{U} statistics (DRGU):
\begin{align}
    U_n^{DR} = \left[ \binom{n}{2}\right]^{-1} \sum_{i,j\in C^n_2 } h^{DR}_{ij},
\end{align}
where, $h_{ij}^{DR} = \frac{z_i(1-z_j)}{2\pi_i(1-\pi_j)}(\varphi(y_{i1} - y_{j0})-g_{ij}) + \frac{z_j(1-z_i)}{2\pi_j(1-\pi_i)}(\varphi(y_{j1} - y_{i0})-g_{ji})+\frac{g_{ij}+g_{ji}}{2}$.
When $\pi$ and $g$ are known, we can show that $E(h_{ij}^{DR})=\delta$, and thus $E(U_n^{DR}) = \delta$ (Appendix~\ref{apx:DRU_Unbias}). Further, variance of $U_n^{DR}$ reaches \textit{semi-parametric bound} (Appendix ~\ref{apx:DRU_Efficiency}), i.e., smallest variance among all unbiased estimator under semi-parametric set-up. In most applications, we don't know $\pi$ and $g$, and need to estimate them via $\hat{\pi}$ and $\hat{g}$. As long as one of $\hat{\pi}$ and $\hat{g}$ is consistent estimator, then the corresponding U statistics $\hat{U}_n^{DR}$, is also consistent, hence doubly robust. 

We can estimate $\pi_i$ and $g_{ij}$ by imposing a linear structure: $\pi(w_i; \beta) = \phi([1,w_i^T]^T \cdot \beta))$, $g(w_i, w_j; \gamma)= \psi([1, w_i^T, w_j^T]^T \cdot \gamma)$, where $\phi()$ and $\psi()$ are link functions. Note that $g()$ is a model on pair of data points and can be considered as simplified Graph Neural Network.

For estimation and inference of the parameters $\theta = (\delta, \beta, \gamma)$, one way is to do it sequentially, i.e., first estimating $\hat{\beta}$ and $\hat{\gamma}$ with the regression models, then calculating $\hat{U}_n^{DR}(\hat{\beta}, \hat{\gamma})$ and corresponding asymptotic variance considering variance from $\hat{\beta}$ and $\hat{\gamma}$. We will leverage U-statistics-based Generalized Estimation Equations (UGEE) \cite{kowalski2008} for joint estimation and inference:
\begin{align}
    \mathbf{U}_n(\theta) = \sum_{i,j\in C^n_2} \mathbf{U}_{n,ij} = \sum_{i,j\in C^n_2} \mathbf{G}_{ij}(\mathbf{h}_{ij}-\mathbf{f}_{ij}) = \mathbf{0},
\end{align}

where, $\mathbf{h}_{ij} = [h_{ij1}, h_{ij2}, h_{ij3}]^T$, 
$\mathbf{f}_{ij} = [f_{ij1}, f_{ij2}, f_{ij3}]^T$, 
$h_{ij1} = \frac{z_i(1-z_j)}{2\pi_i(1-\pi_j)}(\varphi(y_{i1} - y_{j0})-g_{ij}) + \frac{z_j(1-z_i)}{2\pi_j(1-\pi_i)}(\varphi(y_{j1} - y_{i0})-g_{ji})+\frac{g_{ij}+g_{ji}}{2}$,
$h_{ij2} = z_i+z_j$, 
$h_{ij3} = z_i(1-z_j)\varphi(y_{i1} - y_{j0}) + z_j(1-z_i)\varphi(y_{j1} - y_{i0})$,
$f_{ij1} = \delta$, 
$f_{ij2} = \pi_i + \pi_j$, 
$f_{ij3} = \pi_i(1-\pi_j)g_{ij} + \pi_j(1-\pi_i)g_{ji}$, 
$\pi_i = \pi(w_i; \beta)$, 
$g_{ij} = g(w_i, w_j; \gamma)$, and 
$\mathbf{G}_{ij} =  \mathbf{D}_{ij}^T \mathbf{V}^{-1}_{ij}$, 
$\mathbf{D}_{ij} = \frac{\partial \mathbf{f}_{ij}}{\partial \theta}$, 
$\mathbf{V}_{ij} = diag\{Var(h_{ijk}|w_i, w_j)\}$.

\begin{theorem} \label{thm:DR_U}
Let $\mathbf{u}_{i} = E(\mathbf{U}_{n,ij}|y_{i0},y_{i1},z_i, w_i)$, $\Sigma=Var(\mathbf{u}_i)$, $\mathbf{M}_{ij} = \frac{\partial (\mathbf{f}_{ij}-\mathbf{h}_{ij})} {\partial \theta}$, and $\mathbf{B} = E(\mathbf{G}\mathbf{M})$. Let $\hat{\delta}$ be the 1st element in $\hat{\theta}$. 
Then, under mild condition, we have consistency: $\hat{\theta} \rightarrow_p \theta$, and asymptotic normality:
\begin{align}
    \sqrt{n}(\hat{\theta} - \theta) \rightarrow_d N(0, 4B^{-} \Sigma B^{-T}).
\end{align}
Further, as long as one of $\pi$ and $g$ is correctly specified, $\hat{\delta}$ is consistent. When both are correctly specified, $\hat{\delta}$ attains \textbf{semi-parametric efficiency bound}.

\end{theorem}

Proof is provided in Appendix~\ref{apx:DRU_Efficiency} and~\ref{apx:DRU_Asymp}. We can estimate $\theta$ via either one of the following iterative algorithm:
$\theta^{(t+1)} = \theta^{(t)} - (\left. \frac{\partial \mathbf{U}_n(\theta)} {\partial\theta}\right|_{\theta^{(t)}})^{-}\mathbf{U}_n(\theta^{(t)})$, or 
$\theta^{(t+1)} = \theta^{(t)} + (\hat{B}(\theta^{(t)}) )^{-}\mathbf{U}_n(\theta^{(t)})$
where, 
$\hat{B} = \binom{n}{2}^{-1}\sum_{i,j \in C^n_2} \hat{\mathbf{G}}_{ij} \hat{\mathbf{M}}_{ij}$. $\Sigma$ can be estimated empirically from outerproduct of $\hat{\mathbf{u}}_i = \frac{1}{n-1}\sum_{j\neq i} \mathbf{U}_{ij}(\hat{\theta})$, i.e.,
$\hat{\Sigma} = \frac{1}{n}\sum_{i}\hat{\mathbf{u}}_i\hat{\mathbf{u}}_i^T$.

\subsection{DR Generalized U for Longitudinal Data}

Let $y_{it}$ denote the metrics we measures overtime, $z_i$ denote the treatment assignment, and $w_{it}$ denote the variables needs to be adjusted for. We can measure the treatment effect overtime:
$\delta_t = E(\varphi(y_{it1} - y_{it0}))$,
where $y_{it0}$ and $y_{it1}$ are counterfactual responses for $z_i=0$ and $z_i=1$. 
We can construct DR type of multivariate U statistic for the longitudinal data,
\begin{align}
    \mathbf{U}_n^{DR} =  \left[ \binom{n}{2}\right]^{-1} \sum_{i,j\in C^n_2 } \mathbf{h}^{DR}_{ij},
\end{align}
where,
$\mathbf{h}^{DR}_{ij} = [h_{ij1}, \cdots, h_{ijt}, \cdots, h_{ijT}]^T$,
$h_{ijt} = \frac{z_i(1-z_j)}{2\pi_i(1-\pi_j)}(\varphi(y_{it1} - y_{jt0})-g_{ijt}) + \frac{z_j(1-z_i)}{2\pi_j(1-\pi_i)}(\varphi(y_{jt1} - y_{it0})-g_{jit})+\frac{g_{ijt}+g_{jit}}{2}$.

We can estimate $\pi$ and $g$ by,
$E(z_i|\mathbf{w}_i) = \pi(\mathbf{w}_i; \beta) = \phi([1,\mathbf{w}_i^T]^T \cdot \beta))$,
$E(\varphi(y_{it1} - y_{jt0})|w_{it}, w_{jt}) = g(w_{it}, w_{jt}; \gamma_t)= \psi([1, w_{it}^T, w_{jt}^T]^T \cdot \gamma_t)$,
where $\mathbf{w} = [w_1^T, \cdots, w_{t}^T,\cdots w_{T}^T]^T$.

We can estimate the parameters and make inference jointly for $\boldsymbol{\theta} = [\boldsymbol{\delta}^T, \beta^T, \boldsymbol{\gamma}^T]^T$ using UGEE: 
\begin{align}
    \mathbf{U}_n(\theta) = \sum_{i,j\in C^n_2} \mathbf{U}_{n,ij} = \sum_{i,j\in C^n_2} \mathbf{G}_{ij}(\mathbf{h}_{ij}-\mathbf{f}_{ij}) = \mathbf{0},
\end{align}

where,
    $\mathbf{h}_{ij} = [\mathbf{h}_{ij1}^T, h_{ij2}, \mathbf{h}_{ij3}^T]^T$,
    $\mathbf{f}_{ij} = [\mathbf{f}_{ij1}^T, f_{ij2}, \mathbf{f}_{ij3}^T]^T$,
    $\mathbf{h}_{ij1} = \mathbf{h}_{ij}^{DR}$,
    $h_{ij2} = z_i+z_j$,
    $\mathbf{h}_{ij3} = z_i(1-z_j)\boldsymbol{\varphi}_{ij} + z_j(1-z_i)\boldsymbol{\varphi}_{ji}$,
    $\mathbf{f}_{ij1} = \boldsymbol{\delta}$,
    $f_{ij2} = \pi_i + \pi_j$,
    $\mathbf{f}_{ij3} = \pi_i(1-\pi_j)\mathbf{g}_{ij} + \pi_j(1-\pi_i)\mathbf{g}_{ji}$,
    $\pi_i = \pi(\mathbf{w}_i; \beta)$,
    $\mathbf{g}_{ij} = \mathbf{g}(\mathbf{w}_i, \mathbf{w}_j; \boldsymbol{\gamma})$,
and 
    $\mathbf{G}_{ij} =  \mathbf{D}^T_{ij} \mathbf{V}^{-1}_{ij}$,
    $\mathbf{D}_{ij} = \frac{\partial \mathbf{f}_{ij}}{\partial \theta}$,
    $\mathbf{V}_{ij} = A \mathbf{R}(\alpha) A$,
    $A = diag\{\sqrt{Var(h_{ijkt_k}|w_i, w_j)}\}$.
Note here $h_{ij2}$ is scalar and $\mathbf{h}_{ij}$ is a vector of length $2T+1$. 

\begin{corollary}
\label{thm:DR_U_Long}
Let $\mathbf{u}_{i} = E(\mathbf{U}_{n,ij}|\mathbf{y}_{i0},\mathbf{y}_{i1},z_i, \mathbf{w}_i)$, $\Sigma=Var(\mathbf{u}_i)$, $\mathbf{M}_{ij} = \frac{\partial (\mathbf{h}_{ij}-\mathbf{f}_{ij})} {\partial \boldsymbol{\theta}}$, and $\mathbf{B} = E(\mathbf{G}\mathbf{M})$.
Then, under mild condition, we have consistency: $\hat{\boldsymbol{\theta}} \rightarrow_p \boldsymbol{\theta}$, and asymptotic normality: $\sqrt{n}(\hat{\boldsymbol{\theta}} - \boldsymbol{\theta}) \rightarrow_d N(0, 4B^{-} \Sigma B^{-T}).$
\end{corollary}

Estimation and computation of asymptotic variance can be perform in the same manner as Section~\ref{sec:DR_U} for small to medium sample size. For large sample size, the computation burden can grow significantly. We device efficient algorithms for optimization and inference (\textit{Algorithm~\ref{algo:optimization}} and \textit{Algorithm~\ref{algo:inference}}), and provide theoretical support of the algorithms with \textit{Theorem~\ref{thm:decouple}},  \textit{Lemma~\ref{thm:mc_error}} and \textit{Theorem~\ref{thm:anchor-partner}}. (See proof in Appendix~\ref{apx:LD_DRU_Bound})

In most applications, we can reduce number of parameters by imposing some structures on the trajectory ($\gamma_t$ and $\delta_t$), for examples: (i) set the $g_{t}$ to same functional form, i.e, $\gamma_t=\gamma$; (ii) set the $\delta_t$ to be a simple linear form, e.g., $\delta_t = \delta$, or $\delta_t = \delta_1 + \delta_2t$. Our simulation and real application will use these structure. 

\begin{algorithm}[h]
\caption{Mini-batch Fisher Scoring for $\hat{\theta} = (\hat{\delta}, \hat{\beta}, \hat{\gamma})$}
\label{algo:optimization}
\begin{algorithmic}[1]
    \STATE \textbf{Input:} Data $\{(y_i, z_i, w_i)\}_{i=1}^n$, initial parameter $\theta^{(0)}$, step size $\alpha$, batch size $m$, convergence threshold $\varepsilon = c\,n^{-1/2-\varsigma/2}$ for $\varsigma>0$.
    \STATE $t \gets 0$
    \REPEAT
        \STATE Sample $m$ rows without replacement: $S_t = \{i_1, \dots, i_m\}$ from current epoch
        \STATE Form all $\binom{m}{2}$ unordered pairs $\{(i,j) : i < j,\ i,j \in S_t\}$
        \STATE For each pair $i,j$, compute:
        \begin{itemize}
            \item $\mathbf{U}_{ij} = \mathbf{G}_{ij}(\mathbf{h}_{ij} - \mathbf{f}_{ij})$
            \item $\mathbf{B}_{ij} = \mathbf{G}_{ij} \mathbf{M}_{ij}$
        \end{itemize}
        \STATE Estimate score: $\tilde{\mathbf{U}}_t = \frac{2}{m(m-1)} \sum_{i<j} \mathbf{U}_{ij}$
        \STATE Estimate Jacobian: $\tilde{\mathbf{B}}_t = \frac{2}{m(m-1)} \sum_{i<j} \mathbf{B}_{ij}$
        \STATE Update parameter:
        \[
            \theta^{(t+1)} = \theta^{(t)} + \alpha \left(\tilde{\mathbf{B}}_t\right)^{-1} \tilde{\mathbf{U}}_t
        \]
        \STATE $\hat{U}_t = \mathbf{1}_{t>0}[(1-e)\hat{U}_{t-1} + e \tilde{U}_t] + \mathbf{1}_{t=0}\tilde{U}_t$
        \STATE $t \gets t + 1$
    \UNTIL{$\left\| \hat{\mathbf{U}}_t \right\| < \varepsilon$}
    \STATE \textbf{Output:} $\hat{\theta} = \theta^{(t)}$, $\hat{\delta}$ is the first component
\end{algorithmic}
\end{algorithm}

\begin{algorithm}[h]
\caption{Monte Carlo Estimation of $\widehat{Var}(\hat{\theta})$}
\label{algo:inference}
\begin{algorithmic}[1]
    \STATE \textbf{Input:} Data $\{(y_i, z_i, w_i)\}_{i=1}^n$, parameter $\hat{\theta}$ from Fisher scoring, number of anchors $s=c_1n$, number of partners $m=c_2\log n$, de-bias parameter $\alpha \in [0,1]$ 
    \STATE Sample $s$ anchor indices $I=\{i_1,\dots,i_s\}$ uniformly from $\{1,\dots,n\}$
    \FORALL{anchors $i \in I$}
        \STATE Sample $m$ partner indices $J_i \subset \{1,\dots,n\}\setminus\{i\}$ uniformly without replacement
        \STATE For each $j \in J_i$, compute
        $
        \mathbf{u}_{ij} = \mathbf{G}_{ij}(\mathbf{h}_{ij} - \mathbf{f}_{ij})$, 
        $
        \mathbf{B}_{ij} = \mathbf{G}_{ij} \mathbf{M}_{ij}
        $
        \STATE Compute anchor mean: 
        $
        \bar{\mathbf{u}}_i = \frac{1}{m}\sum_{j\in J_i}\mathbf{u}_{ij}
        $
        \STATE Compute within-anchor covariance:
        \[
        \hat{\Sigma}_i^{\,\text{within}} = \frac{1}{m-1}\sum_{j\in J_i} (\mathbf{u}_{ij}-\bar{\mathbf{u}}_i)(\mathbf{u}_{ij}-\bar{\mathbf{u}}_i)^\top
        \]
    \ENDFOR
    \STATE Estimate Jacobian:
    $
    \hat{B} = \frac{1}{s}\sum_{i\in I}\frac{1}{m}\sum_{j\in J_i}\mathbf{B}_{ij}
    $
    \STATE Estimate variance component:
    \[
    \hat{\Sigma} = \frac{1}{s}\sum_{i\in I} \bar{\mathbf{u}}_i \bar{\mathbf{u}}_i^\top
    \;-\alpha\;\Big(\tfrac{1}{m}-\tfrac{1}{n-1}\Big)\cdot \frac{1}{s}\sum_{i\in I}\hat{\Sigma}_i^{\,\text{within}}
    \]
    \STATE \textbf{Output:}
    \[
    \widehat{Var}(\hat{\theta}) \;=\; \frac{4}{n}\,\hat{B}^{-}\,\hat{\Sigma}\,\hat{B}^{-T}
    \]
\end{algorithmic}
\end{algorithm}

\begin{theorem}[Decoupling of Optimization and Inference]\label{thm:decouple}
Assume the estimating equation
\[
\bar U_n(\theta)=  \frac{1}{\binom{n}{2}}\sum_{i,j\in C^n_2} U_{n,ij}(\theta) = 0
\]
is solved by a numerical algorithm producing $\hat\theta$ such that
\[
\|\bar U_n(\hat\theta)\| = o_p\bigl(n^{-1/2}\bigr).
\]
Then, one has
$
\sqrt{n}\,(\hat\theta - \theta) \rightarrow_d
N\bigl(0,\;4\,B^{-}\,\Sigma\,B^{-T}\bigr).
$
In particular, the small algorithmic error does not affect the first‐order asymptotic distribution.  
\end{theorem}

\begin{lemma}[Monte Carlo Error Bound]\label{thm:mc_error}
Let
$U_n^v =\binom n2^{\!-1}\sum_{i<j}v(o_i,o_j),$
with symmetric, sub‐Gaussian kernel $v$ (proxy variance $\sigma^2$).  Form the Monte Carlo estimator
$\hat U_k =\frac1k\sum_{(i,j)\in C_k}v(o_i,o_j),$
where $k$ pairs are sampled uniformly without replacement from the $\binom n2$ possible, and let the average overlap factor be \(\Delta = O(k/n)\).  Then for any \(\epsilon>0\) and \(\eta\in(0,1)\),
\[
P\bigl(|\hat U_k - E[\hat U_k]|>\epsilon\bigr)
\;\le\;2\exp\!\Bigl(-\tfrac{k\,\epsilon^2}{2\,\sigma^2\,(1+\Delta)}\Bigr),
\]
and hence with effective sample size \(\tilde k = k/(1+\Delta)\),
\[
|\hat U_k - E[U_n^v]|\;\le\;\sqrt{\frac{2\,\sigma^2}{\tilde k}\,\log\!\Bigl(\tfrac2\eta\Bigr)}
\quad\text{w.p. }1-\eta.
\]
In particular,
\[
\hat U_k - E[U_n^v]
\;=\;O_p\Bigl(\sqrt{\tfrac1k+\tfrac1n}\Bigr).
\]
\end{lemma}

\begin{theorem}[MC Error for Anchor--Partner Estimator]
\label{thm:anchor-partner}
Under sub-Gaussian tails, the anchor--partner scheme with $s$ anchors and $m$ partners per anchor yields
\[
\hat B - B = O_p(s^{-1/2}) + O_p((sm)^{-1/2}) + O_p(n^{-1/2}), 
\]
\[
\hat\Sigma - \Sigma = O_p(s^{-1/2}) + O_p(n^{-1/2}) + O(m^{-1}) + O(n^{-1}).
\]
With de-biasing, the $O(m^{-1})$ term vanishes. 
By a first-order expansion, the variance estimator 
$\hat V = 4\,\hat B^{-1}\hat\Sigma \hat B^{-T}$ inherits the same rate. 
In particular, choosing $s=O(n)$ gives 
\[
\hat V - V = O_p(n^{-1/2}),
\]
matching the efficiency of the full $O(n^2)$ estimator.
\end{theorem}

\section{Experiments and Results}
\label{sec:SimuApp}

\subsection{Simulation Studies}
We perform comprehensive simulation studies to evaluation performance of the proposed methods. Due to space limitation, we summarize  and highlight the results here.
\subsubsection{Regression Adjustment} When there is no confounding, both t-test and RA can control type I error, while RA has higher power than t-test. Under confounding, t-test can't control type I error while RA can control type I error. (Appendix ~\ref{apx:RA_Simu})
\subsubsection{GEE} We simulate confounding effect and repeated measurement. Both snapshot regression and GEE can control type I error under confounding, while GEE has higher power. (Appendix ~\ref{apx:GEE_Simu})
\subsubsection{ZTU} For heavy tailed distribution with $50\%$ of zeros, Zero-trimmed U has higher power than standard Mann Whitney U most of the time and standard U has higher power than t-test. All three methods can control type I error for zero inflated heavy tail data. (Appendix ~\ref{apx:MWU_Simu})

\begin{table}[h]
\centering
\caption{Power Comparison for Heavy Tailed Distributions with Equal Zero-Inflation (50\%)}
\label{tab:power_comparison_wide}
\begin{tabular}{c|ccc|ccc}
\hline
Effect Size & \multicolumn{3}{c|}{Positive Cauchy} & \multicolumn{3}{c}{LogNormal } \\
 & ZTU & MWU & t & ZTU & MWU & t \\
\hline
0.25 & \textbf{0.079} & 0.065 & 0.011 & 0.044 & 0.044 & 0.009 \\
0.50 & \textbf{0.165} & 0.094 & 0.026 & \textbf{0.067} & 0.059 & 0.004 \\
0.75 & \textbf{0.339} & 0.166 & 0.031 & \textbf{0.090} & 0.067 & 0.007 \\
1.00 & \textbf{0.555} & 0.262 & 0.048 & \textbf{0.138} & 0.082 & 0.011 \\
\hline
\end{tabular}
\end{table}

\subsubsection{Doubly Robust Generalized U} We simulate confounding effect with heavy tailed distribution. We compare Type I error rates and power of correctly specified \texttt{DRGU}, correctly specified linear regression \texttt{OLS}, and Wilcoxon rank sum test \texttt{U} (which does not account for confounding covariates). To probe double robustness, we set up \texttt{misDRGU} as misspecifying the quadratic outcome propensity score model with a linear mean model, while the outcome model in \texttt{misDRGU} is specified correctly. (Appendix ~\ref{apx:DRU_Simu})

\begin{table}[H]
\centering
\caption{Power of DRGU Adjusting for Confounding Effect}
\label{tab:power_ugee}
\begin{tabular}{lc|cccc}
\hline
Distribution & Sample size & DRGU & misDRGU & OLS & U\\
\hline
\multirow{2}{*}{Normal} & 200 & 0.750 &  0.585 & \textbf{0.940} & 0.299 \\
& 50 & \textbf{0.135} & 0.085 & \textbf{0.135} & 0.035 \\
\hline
\multirow{2}{*}{LogNormal} & 200 & \textbf{0.610} &  0.515 & 0.435 & 0.235 \\
& 50 & \textbf{0.260} & 0.210 & 0.190 & 0.110 \\
\hline
\multirow{2}{*}{Cauchy} & 200 & \textbf{0.660} & 0.580 & 0.435 & 0.310\\
& 50 & \textbf{0.265} & 0.180 & 0.165 & 0.130 \\
\hline
\end{tabular}
\end{table}



\subsubsection{Longitudinal DRGU} We compare three models \texttt{longDRGU}, \texttt{DRGU} using the last timepoint data snapshot, and \texttt{GEE}. The time-varying covariates highlight the strength of using longitudinal method compared to snapshot analysis. (Appendix ~\ref{apx:LDRU_Simu})

\begin{table}[h]
\centering
\caption{Power of DRGU for Longitudinal data}
\label{tab:power_lugee}
\begin{tabular}{lc|ccc}
\hline
Distribution & Sample size & Long DRGU & DRGU & GEE\\
\hline
\multirow{2}{*}{Normal} & 200 & 0.85 & 0.88 & \textbf{0.92}\\
& 50 & 0.52 & 0.39 & \textbf{0.75}\\
\hline
\multirow{2}{*}{LogNormal} & 200 & \textbf{0.85} & 0.78 & 0.68\\
& 50 & \textbf{0.37} & 0.30 & 0.33\\
\hline
\multirow{2}{*}{Cauchy} & 200 & \textbf{0.83} & 0.76 & 0.66\\
& 50 & \textbf{0.38} & 0.32 & 0.29 \\
\hline
\end{tabular}
\end{table}

\subsection{A/B Tests at LinkedIn}

\subsubsection{Email Marketing} We conducted an user level A/B test comparing our legacy email marketing system against a newer version based on Neural Bandit. We measured the downstream impact on conversion value, a proprietary metric measuring the value of conversions. The conversion value presented characteristic of extreme zero inflation (>95\%) and heavy tailed (among the converted). Using the \textit{Zero-trimmed U} test, we detect a statistically significant lift (+0.94\%) in overall conversion value (p-value<0.001). By constast, the $t$-test is not able to detect a significant effect on the conversion value metric (p-value = 0.249). (Appendix ~\ref{apx:Email_AB})

\subsubsection{Targeting in Feed} We conducted a user level A/B test to evaluate impact of a new algorithm for marketing on a particular slot in Feed. We faced two challenges: (i) selection bias in ad impression allocation that favored the control system, so we need to adjust for impressions as a cost and compare ROI between control and treatment; (ii) imbalance in baseline covariates due to limited campaign and participant selection (Appendix Table \ref{tab:lol}). We addressed both issues via \textit{Regression Adjustment} to estimate ROI lift while controlling for imbalanced covariates, detecting a 1.84\% lift in conversions per impression (95\% CI: [1.64\%, 2.05\%], p<0.001). By contrast, a simple t‑test found no significant difference in conversion (p=0.154). (Appendix ~\ref{apx:LoL_AB})

\subsubsection{Paid Search Campaigns} We ran a 28-day campaign level A/B test on 3rd-party paid-search campaigns (32 control vs. 32 treatment), measuring conversion value net of cost. 

To address the small-sample limitation, we fit a \textit{GEE} model to take advantage of repeated measurement over 28 days, yielding a near-significant effect on ROI (p=0.051) v.s p=0.184 from last day snapshot regression analysis. A 28-day pre-launch AA validation using the same GEE showed no effect (p=0.82), further validating experiment and results. 

\begin{figure}[!htb]
  \centering
    \includegraphics[width=0.48\textwidth]{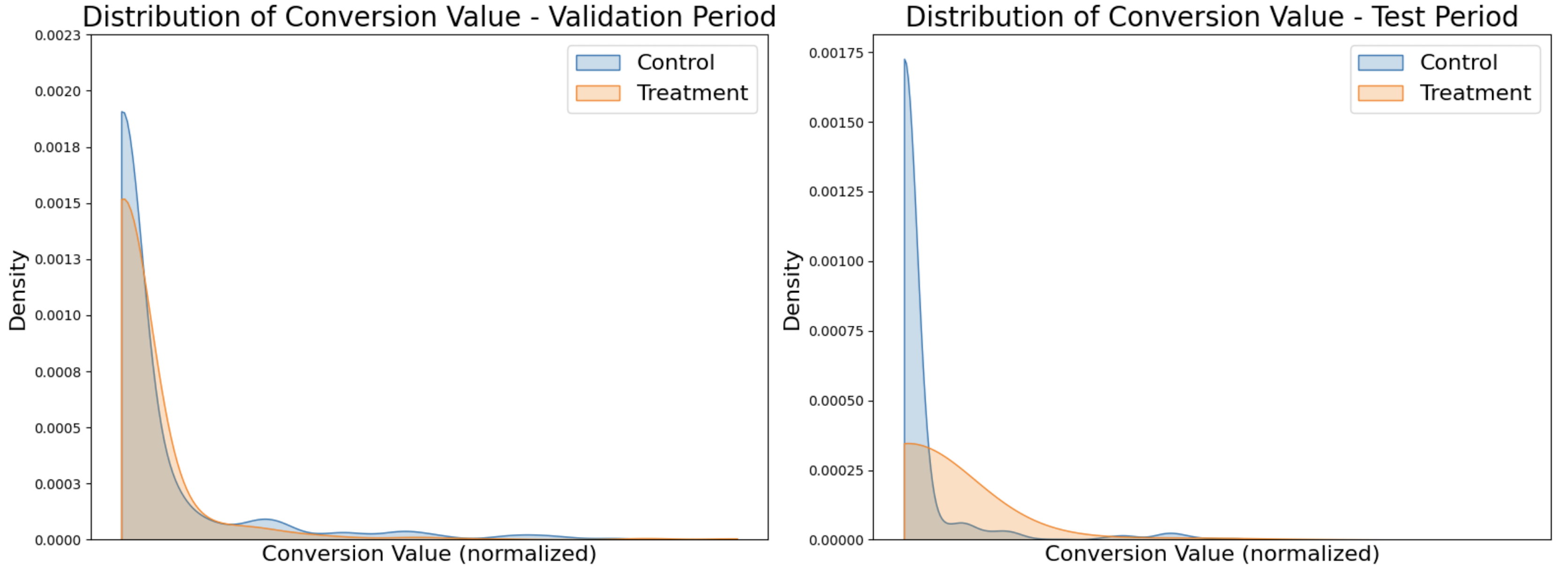}
  \caption{Distribution of Conversion Values from the Validation \& Test Period}
  \label{fig:distribution_gee}
\end{figure}

Observing that the distribution of the conversion value exhibit heavy tail characteristics, we further performed statistical testing using longitudinal \textit{Doubly Robust U}, assuming compound symmetric correlation structure for $R(\alpha)$. We were able to attain statistical significant result with $\hat{P}(y_1>y_0)=0.54$ and $p$-value=0.045. (Appendix ~\ref{apx:Paid_AB})



\section{Discussion}
\label{sec:Discn} 
We provide discussion for general approaches for large sample size (e.g., member level AB test at global scale) as well as various consideration of practical implementation in \textit{Appendix~\ref{apx:Algo_Large}}. 

We further highlight the key contributions on theoretical development and discuss the comparison with existing approaches.

\subsection{Application Innovation}
Although RA, GEE, and the Mann–Whitney U test are established statistical methodologies, their applications to A/B testing are not common practice. This is mainly due to four reasons:
(i) A/B tests in internet companies generally involve huge sample sizes, and efficiency is often not the primary concern;
(ii) for huge sample sizes, RA, GEE, and Mann–Whitney U lack computationally efficient algorithms;
(iii) the primary metrics in A/B tests are typically binary or count data (e.g., impressions or conversions), so there is little perceived need for distribution-robust tests like the Mann–Whitney U;
(iv) evaluation of multiple metrics is often conducted heuristically—e.g., requiring nonsignificance on guardrail metrics and significance on primary metrics, or making ad hoc trade-offs between them.

In A/B tests for business scenarios, the above four reasons vanish:
(i) sample sizes are limited because each A/B test incurs business cost, so using more powerful statistical tests (e.g., covariate adjustment) and increasing effective sample size (e.g., repeated measurement) is very important;
(ii) in many cases, sample sizes are moderate, so computational burden is less of a concern;
(iii) the primary metrics are often revenue, which follows a non-Gaussian distribution, calling for nonparametric tests such as the Mann–Whitney test;
(iv) a principled way of performing ROI trade-offs is needed, and covariate adjustment can measure revenue net of cost. Moreover, when revenue- or value-based primary metrics are used, they are almost always associated with zero inflation and heavy-tail distributions. In this situation, we can use Zero-Trimmed U.

In fact, we argue that these approaches can be applied generally to all A/B tests. Primary metrics can be revenue-based even for engagement-related platforms (e.g., assigning a proxy long-term value to any impression or conversion). Also, there are implicit and explicit costs for any A/B test (e.g., latency can be modeled as a cost to the user). We'll then need robust statistics to address the irregular distribution on proxy value and covariate adjustment for ROI consideration.

For general applicability, we provide ways to efficiently perform the above tests for extremely large sample sizes. RA and GEE are based on estimating equations, and we can use mini-batch Fisher scoring to solve those equations and then calculate variance from the full sample using asymptotic results. Mann–Whitney U and Zero-Trimmed U can be calculated efficiently using fast ranking algorithms, and the variance of the test statistic can be calculated from the asymptotic distribution easily.

\subsection{Theoretical Development}

\subsubsection{Efficiency of RA, GEE and MWU} We provide analytical results for insights into where efficiency gains arise for RA, GEE, and the Mann–Whitney U test:

1) For RA, when there is confounding, relative efficiency over the t-test (measured by MSE) is dominated by the bias term, since the t-test yields a biased estimate of the treatment effect. When there is no confounding, RA’s efficiency gain over the t-test arises from variance reduction due to covariate adjustment. The insight, then, is to find covariates that (i) satisfy non-confounding (i.e., are independent of treatment assignment) and (ii) explain variance in the response. This also explains the efficiency gains of related CUPED-type methods.

2) For GEE, we show that efficiency gains over snapshot come from using repeated measurements, and we derive the exact formula for relative efficiency, revealing its dependence on the correlations structure among repeated measurements. When repeated measurements are fully independent, relative efficiency is highest, T times that of snapshot regression. When they are perfectly correlated, GEE and snapshot regression share the same efficiency.

3) For the Mann–Whitney U test, we compute relative efficiency over the t-test on several example distributions, illustrating near-1 efficiency for Gaussian data and higher efficiency for heavy-tailed distributions.

\subsubsection{ZTU} We detail the asymptotics for Zero-Trimmed U, building on existing works from biostatistics field \cite{hallstrom2010,wang2023}. Moreover, we provide a rigorous treatment of Pitman efficiency under compound hypothesis testing in \textbf{\textit{Theorem~\ref{thm:Zero_Eff}}}. Pitman efficiency is given for both (i) Zero-Trimmed U versus Mann–Whitney U with adjusted variance and (ii) Zero-Trimmed U versus Mann–Whitney U with standard (unadjusted) variance.

1) As shown in Figure~\ref{fig:rphi}, the efficiency of Zero-Trimmed U versus Mann–Whitney U with adjusted variance is not always greater than one; it depends on both the direction $\phi$ and the zero proportion $1-p$. When the direction is more on the $d$ component (a location shift among positive values), Zero-Trimmed U has higher power. When the direction focuses on the $m$ component (the zero-proportion difference), Mann–Whitney U with adjusted variance is more efficient, though still close to one. In fact, if $\sin\phi=1$ (purely on $d$), Zero-Trimmed U always has higher power; if $\sin\phi=0$ (purely on $m$), Mann–Whitney U with adjusted variance always has higher power.

2) The efficiency of Zero-Trimmed U versus Mann–Whitney U with standard (unadjusted) variance, however, is mostly greater than one. As shown in Figure~\ref{fig:rphi-o}, the dominance of Zero-Trimmed U is particularly significant (i.e., $r>5$) for high sparsity of positive values ($p$ close to zero); and when there is a substantial proportion of zeros (e.g., $p=0.5$), its advantage is robust to direction (i.e., $\phi$) of the compound hypothesis.

\subsubsection{DRGU} Building on existing works from causal inference Mann-Whitney U in biostatistics field\cite{wu2014,chen2014,vermeulen2015increasing,chen2024,yin2024highly}, we propose a novel \textit{doubly robust \textbf{generalized U}} to address ROI, repeated measurement and distribution robustness all in one framework. We provide the asymptotic results in \textit{\textbf{Theorem~\ref{thm:DR_U} and Corollary~\ref{thm:DR_U_Long}}} with detailed derivations in Appendix~\ref{apx:DRU_Efficiency} and~\ref{apx:DRU_Asymp}. Besides the fact that the application of doubly robust U is completely new for A/B test in business setting (and generally in tech field), we also highlight the key theoretical innovations of DRGU on top of existing approaches:

1) The doubly robust generalized U can adopt any monotonic “kernel” $\varphi$ to form a U statistic to measure the \textit{directional} treatment effect $E(\varphi)$ of a customized definition in an A/B test. When $\varphi$ is the identity function, it reduces to the common doubly robust version of the “mean difference” treatment effect. When $\varphi$ is the indicator function, it is equivalent to the doubly robust version of Mann–Whitney U. There are two key requirements for the kernel $\varphi$: (i) \textit{finite second moment} ensures distributional robustness, i.e. $\mathbb{E}[\varphi^2]<\infty$, a condition the identity kernel (mean-difference) cannot satisfy; (ii) \textit{monotonicity} guarantees that $\varphi$ preserves the test’s directional nature, so that any directional (location) shift in outcomes yields a consistent change in the statistic.

2) We provide a detailed UGEE formulation on joint estimation of both the target parameter $\delta$ and nuisance parameters (i.e., $\beta$, $\gamma$). UGEE is an extension of GEE to pair-wise estimating equations, and readers can refer to \cite{kowalski2008} for a comprehensive treatment of UGEE. Our UGEE formulation is built on top of the formulations from \cite{wu2014, chen2024}. There are three important distinctions: (i) our UGEE is built on a generalized kernel $\varphi$; (ii) we treat $h_{ij3}$, the estimating equation for the “observed” treatment effect, by multiplying the pairwise “missing” probability $z_i(1-z_j)$ with the potential pairwise outcome $\varphi(y_{i1}-y_{j0})$, whereas the formulation in \cite{chen2024} omits the "missing" probability; (iii) we provide the UGEE formulation for longitudinal data, detailing the structure of the propensity model and pairwise regression model for the doubly robust estimator, and the functional forms for different types of longitudinal effects.

3) Besides the asymptotic normality result, we prove that when $\pi$ and $g$ are known, the corresponding estimator attains the semi-parametric efficiency bound, i.e., the proposed doubly robust generalized U has the smallest variance (most powerful) among all regular estimators of the corresponding treatment effect. We further prove that even when $\pi$ and $g$ are unknown, as long as they are correctly specified, the doubly robust generalized U from our UGEE still attains the semi-parametric efficiency bound. This result is stated in Theorem~\ref{thm:DR_U}, which provides the theoretical foundation for its superior performance in simulation and real A/B analysis.

4) We provide computationally efficient algorithms for the proposed doubly robust generalized U on extremely large datasets (e.g., on the order of $10^8$ rows). Basically, the algorithm decouples the optimization procedure that performs the point estimation of $\theta$ and the inference procedure that estimates the asymptotic variance of $\hat{\theta}$. The optimization is driven by mini-batch Fisher scoring on paired data and can be implemented easily with existing automatic differentiation libraries (e.g., JAX, PyTorch, TensorFlow). The inference is driven by Monte Carlo integration for the expectation of variance estimate, where we reduce the computational burden from $O(n^2)$ to $O(n)$ (a huge reduction when $n$ is extremely large) without losing asymptotic efficiency. We provide rigorous theoretical support for the algorithm, on both the decoupling and error bounds, in Appendix~\ref{apx:Algo_Large}. Basically: (i) as long as the mini-batch Fisher scoring algorithm attains error $o_p(n^{-\frac{1}{2}})$, this error is negligible (compared with “perfect” optimization) and thus we can decouple optimization and inference; (ii) as long as the Monte Carlo integration processes a sample of size $O(n^{1+\epsilon})$, the Monte Carlo errors are negligible and we attain the same asymptotic efficiency as using the full $O(n^2)$ pairs.

Besides the application innovation and theoretical development, we also share the JAX\cite{jax2020github} based implementation of UGEE for doubly robust generalized U, as well as code for all simulations, including RA, GEE, Zero-Trimmed U and DRGU, in our Open Sourced Package \textit{robustInfer} (\href{https://github.com/linkedin/robustInfer}{https://github.com/linkedin/robustInfer}). 

\section{Conclusion}
\label{sec:conclusion}
To conclude, we recommended a series of existing and new statistical methods for A/B tests in business settings, with systematic theoretical development and comprehensive empirical evaluations on statistical efficiency. Though these methods are discussed or proposed for business settings, the learning is broadly useful and applicable to general A/B tests and traditional experiments.


\bibliographystyle{ACM-Reference-Format}
\bibliography{beyond_basic_ab}

\appendix
\label{sec:apx}

\section{Algorithm for Large Sample Size}
\label{apx:Algo_Large}
Although the methods proposed are mainly for business scenario, where sample size is  often small to medium, there are business use-case where sample size is large (e.g., large scale marketing campaigns where user level data is available). Moreover, for broader applicability of the methodologies, we need to consider general AB tests in tech where sample size can be at magnitude of million to billion.

For \textit{Mann-Whitney U} and \textit{Zero-trimmed U}, we can leverage fast ranking algorithm to compute $W$ or $W'$. The variance calculation is straightforward using equation in Section ~\ref{sec:NonGausDist}. 

As for \textit{Regression Adjustment}, \textit{GEE} and \textit{DRGU}, they are all based on estimating equations. DRGU have additional layer complexity as its computation is over pairs of observations. We provide efficient algorithm for DRGU in this section. The algorithm for RA and GEE should follows trivially. 

\subsection{Large Data Estimation and Inference for DR Generalized U}
\label{apx:LD_DRU}

The high level idea is to decouple optimization (solving UGEE) and inference (estimation of variance), and use efficient algorithm for both steps:
\begin{enumerate}
    \item \textit{Optimization} : We obtain $\hat{\theta}$ by stochastic Fisher scoring with mini‑batches until $||\bar{U}_n||<cn^{-\frac{1}{2}(1+\varsigma)}$ (i.e., $||\bar{U}_n||=o_p(n^{-\frac{1}{2}})$).
    \item \textit{Inference} : We estimate $B=E(GM)$ and $\Sigma=E(\mathbf{u}\mathbf{u}^T)$ with Monte Carlo integration from subsample of pairs, and calculate $\widehat{Var}(\hat{\theta})=\frac{4\hat{B}^-\hat{\Sigma}\hat{B}^{-T}}{n}$.
\end{enumerate}
Details are described in Algorithm~\ref{algo:optimization} and Algorithm~\ref{algo:inference}. 

\textit{Remarks}:
\begin{itemize}
    \item For Algorithm~\ref{algo:optimization}, we can use sample by pairs instead of by rows. Both give consistent estimate of the parameter, i.e., $\hat{\theta}\rightarrow_p\theta$. There are trade-off on multiple aspects: (i) sampling by pair gives clean guarantee on unbiasedness while sampling by row can be biased (though consistent) due to missing on intra-batch pairs; (ii) sampling by row is easier to implement and can use GPU more efficiently while sampling by pair needs to generate all pairs beforehand or implement reservoir sampling (or hashing tricks) for extreamly large data. For both approaches, stratified sampling should be used for highly imbalanced data. 

    \item For Algorithm~\ref{algo:inference}, the Monte Carlo error is controlled by the number of anchors $s$ and the number of partners $m$. 
    Choosing $s$ proportional to $n$ ensures root-$n$ accuracy, while setting $m$ to grow slowly (e.g., $m=c_2 \log n$) is sufficient for stable within-anchor covariance estimates. 
    After de-biasing, the asymptotic error rate no longer depends on $m$, so $m$ only affects finite-sample stability. 
    For example, with data size $10^8$, setting $s \approx 10^7$ and $m \approx 100$ already yields practical inference with $O_p(n^{-1/2})$ accuracy. 
    By contrast, there is no need to compute all $O(n^2)$ pairs. 
    Note that for standard regression and GEE settings, one can compute variance directly on the full sample without Monte Carlo integration.
    \item The working correlation matrix $R(\alpha)$ can be estimated in an outer loop around the $\theta$-updates, e.g., by alternating between updating $\theta$ using Fisher scoring and re-estimating $\alpha$ based on current residuals: (i) A good initial value for $\alpha$ is typically $\alpha^{(0)} = 0$, corresponding to the independence working correlation, which ensures consistency of $\hat{\theta}$ even if $R(\alpha)$ is misspecified; (ii) $\alpha$ can be re-estimated every $K$ steps of the inner Fisher scoring loop. This avoids excessive overhead from updating $\alpha$ too frequently. (iii) Re-estimation of $\alpha$ can stop once its updates become small or after a fixed number of outer iterations. Typically, only a few updates (e.g., $3\sim5$) are sufficient in practice.
\end{itemize}

\subsection{Algorithms Decoupling and Error Bounds}
\label{apx:LD_DRU_Bound}

\subsubsection{Algorithms Decoupling}
To see why we can decuple the optimization and inference (i.e., Algorithm~\ref{algo:optimization} and Algorithm~\ref{algo:inference}), observe that 

\begin{align*}
    \sqrt{n}\bar{U}_n(\hat{\theta}) &= \sqrt{n}\bar{U}_n(\theta)+ \frac{\partial \bar{U}_n}{\partial \theta}\sqrt{n}(\hat{\theta}-\theta)+o_p(1)\\
    \sqrt{n}(\hat{\theta}-\theta) &= - (\frac{\partial \bar{U}_n}{\partial \theta})^- \sqrt{n} \bar{U}_n(\theta) + (\frac{\partial \bar{U}_n}{\partial \theta})^- \sqrt{n} \bar{U}_n(\hat{\theta}) + o_p(1)
\end{align*}

The second term measure the "error" when the estimating equation is not exactly solved, i.e., algorithm error. The first term measures the sampling variations. When the fisher scoring algorithm error is small, $||\bar{U}_n||=o_p(n^{-\frac{1}{2}})$, we know 
\[(\frac{\partial \bar{U}_n}{\partial \theta})^- \sqrt{n} \bar{U}_n(\hat{\theta}) = O_p(1)\sqrt{n}o_p(n^{-\frac{1}{2}}) = o_p(1)\]
and thus
\[\sqrt{n}(\hat{\theta}-\theta) = - (\frac{\partial \bar{U}_n}{\partial \theta})^- \sqrt{n} \bar{U}_n(\theta) + o_p(1) \rightarrow_d N(0, 4B^-\Sigma B^{-T}).\]
We state the above results in Theorem~\ref{thm:decouple}

\subsubsection{MC Error Bound for U statistics}
Without loss of generality, let's assume the symmetric kernel $ v(o_i, o_j) \in \mathbb{R} $ is sub-Gaussian with proxy variance $\sigma^2$.

We compute a Monte Carlo approximation
$
\hat{U}_k = \frac{1}{k} \sum_{(i,j)\in C_k} v(o_i, o_j)
$
by sampling $ k $ unordered pairs from the full set of $ \binom{n}{2} $ possible pairs.
Due to overlapping indices among pairs, the kernel evaluations are not fully independent. Observe that, for all sampled pair, the expected total number of overlapping pairs are $O(\frac{k^2}{n})$. Then, for each sampled pair, the number of overlapping pairs is 
\[\Delta = O(k/n), \] 
and hence $Var(\hat{U}_k) = \frac{1}{k^2}\sum_{l\in C_k}Var(v_l)+\frac{1}{k(k-1)}\sum_{l\neq l'}Cov(v_l,v_{l'})=\frac{\sigma^2}{k}+O(\frac{1}{n})C =\frac{\sigma^2}{k}(1+\Delta)$, provided that $Cov(v_l,v_{l'})\leq C$.

Using Bernstein-type inequalities \cite{boucheron2013concentration} adapted for $Var(\hat{U}_k)=\frac{\sigma^2}{k}(1+\Delta)$, the Monte Carlo average satisfies
\[
P\left( \left| \hat{U}_k - E[\hat{U}_k] \right| > \epsilon \right)
\leq 2 \exp\left( - \frac{ k \epsilon^2 }{ 2\sigma^2 (1 + \Delta) } \right)
\]
This introduces an adjustment factor $1 + \Delta$ into the denominator, reflecting variance inflation due to overlap between sampled pairs.

To achieve a target error $\epsilon$ with confidence level $1 - \eta$, we can set $ 2 \exp\left( - \frac{ k \epsilon^2 }{ 2\sigma^2 (1 + \Delta) } \right) \leq \eta$. Solving this w.r.t "effective sample size" $\tilde{k}=k/(1+\Delta)$, we have 
\[
\tilde{k}\geq \frac{2\sigma^2}{\epsilon^2} \log(\frac{2}{\eta}).
\]

Equivalently, with high probability $1-\eta$, the finite sample error bound is:
\[\left| \hat{U}_k - \mathbb{E}[U_n] \right| \leq \sqrt{\frac{2\sigma^2}{\tilde{k}}\log(\frac{2}{\eta})}\]

The bound implies that the effective asymptotic convergence rate is
\[
\hat{U}_k - \mathbb{E}[U_n]  = O_p\left( \sqrt{ \frac{1}{\tilde{k}} } \right)
= O_p\left( \sqrt{ \frac{1}{k} + \frac{1}{n} } \right) = O_p\left (\sqrt{\frac{1}{n}(1+\frac{n}{k})}\right )
\]

We state the above results in Lemma~\ref{thm:mc_error}.

\subsubsection{MC Error Bound for Anchor--Partner Estimator}

We now analyze the anchor--partner scheme with $s$ anchors and $m$ partners per anchor.

\paragraph{Jacobian matrix $B$.}
For each anchor $i$ and partner $j\in J_i$, we compute 
$\mathbf{B}_{ij} = \mathbf{G}_{ij}\mathbf{M}_{ij}$ and form
\[
\hat B = \frac{1}{sm}\sum_{i=1}^s\sum_{j\in J_i}\mathbf{B}_{ij}.
\]
Then $\mathbb{E}[\hat B]=B$. Because all $m$ partners within an anchor share the same $O_i$, the variance has two components:
\[
Var(\hat B) \;=\; \frac{\Omega_B}{s} \;+\; \frac{\tau_B^2}{sm} \;+\; O\!\left(\frac{1}{n}\right),
\]
where $\Omega_B = Var(\mathbb{E}[\mathbf{B}_{ij}\mid O_i])$ (between anchors) and $\tau_B^2=\mathbb{E}[Var(\mathbf{B}_{ij}\mid O_i)]$ (within anchors).  
Thus
\[
\hat B - B \;=\; O_p\!\Big(\tfrac{1}{\sqrt{s}}\Big) + O_p\!\Big(\tfrac{1}{\sqrt{sm}}\Big) + O_p\!\Big(\tfrac{1}{\sqrt{n}}\Big).
\]
The leading term is usually $O_p(1/\sqrt{s})$, reflecting anchor reuse; increasing $m$ mainly reduces the smaller within-anchor term.

\paragraph{Variance component $\Sigma$.}
In the anchor--partner scheme, each anchor $i$ has a Monte Carlo row mean
$\bar u_i = \frac{1}{m}\sum_{j\in J_i} u_{ij}$,
and we estimate 
\[
\hat\Sigma = \frac{1}{s}\sum_{i\in I} \bar u_i \bar u_i^\top.
\]

Because each $\bar u_i$ averages only $m$ partners from $n-1$, we have
\[
E[\bar u_i \bar u_i^\top] 
= \psi(O_i)\psi(O_i)^\top + \frac{1}{m}\,Var(u(O_i,O')\mid O_i) 
+ O\!\left(\tfrac{1}{n}\right),
\]
and thus the bias term:
\[
E[\hat\Sigma] - \Sigma = O\!\left(\tfrac{1}{m}\right) + O\!\left(\tfrac{1}{n}\right).
\]

Across anchors, the $\bar u_i$ are sub-Gaussian and nearly independent. Different anchors share $O(m^2/n)$ partners in expectation, which induces correlations of order $O(1/n)$. Concentration for weakly dependent arrays gives the fluctuation term:
\[
|\hat\Sigma - E[\hat\Sigma]| = O_p\!\left(\sqrt{\tfrac{1}{s}}\right) + O_p\!\left(\tfrac{1}{\sqrt{n}}\right).
\]

Combining bias and fluctuation term,
\[
\hat\Sigma - \Sigma = O_p\!\left(s^{-1/2}\right) + O_p\!\left(n^{-1/2}\right) + O(m^{-1}) + O(n^{-1}).
\]

Since the $O(m^{-1})$ bias can dominate, we apply a de-bias step by subtracting an $\alpha$-fraction of the within-anchor covariance. Setting $\alpha=1$ removes the full $O(m^{-1})$ term, while smaller $\alpha\in[0,1]$ can be used in practice to enforce positive semi-definiteness. After de-biasing,
\[
\hat\Sigma - \Sigma = O_p\!\left(s^{-1/2}\right) + O_p\!\left(n^{-1/2}\right) + O(n^{-1}).
\]

\paragraph{Summary.}
For the anchor--partner estimator:
\[
\hat B - B = O_p(s^{-1/2}) + O_p((sm)^{-1/2}) + O_p(n^{-1/2}), \]
\[
\hat\Sigma - \Sigma = O_p(s^{-1/2}) + O_p(n^{-1/2}) + O(n^{-1}),
\]
and, by a first-order expansion, the variance estimator $\hat V = 4\,\hat B^{-1}\hat\Sigma \hat B^{-T}$ inherits the same error rate:
\[
\hat{V} - V
= O_p(s^{-1/2}) + O_p(n^{-1/2}) + O(n^{-1}).
\]
After de-biasing, the $O(m^{-1})$ term vanishes, so the choice of $m$ only affects finite-sample stability but not the asymptotic rate. In particular, choosing $s=O(n)$ yields 
$\hat V - V = O_p(n^{-1/2})$, achieving the same efficiency as the full $O(n^2)$ estimator at substantially lower cost.

\section{Efficiency of repeated measurement analysis}

\subsection{Efficiency of GEE over snapshot regression}
\label{apx:GEE_efficiency}
We derive the Asymptotic Efficiency of GEE over snapshot Regression on repeated measurement linear model, shown in Section~\ref{sec:LA}. 

We can write the underlying linear model as
$y_i = X_i\theta + \epsilon_i$,
where $X_i=vx_i^T$, $v=[1,\cdots,1,\cdots,1]^T$. 

For GEE of above model, we know 
$D_i=\frac{\partial \mu_i}{\partial \theta} = X_i$, $V_i^{-1} = \frac{1}{\sigma^2}R^{-1}$,  
$\hat{B}= \frac{1}{n}\sum_i D_i^TV_i^{-1}D_i$,
$Var(\hat{\theta}_{gee}) = \frac{\hat{B}^{-T}\hat{\Sigma}_u\hat{B}^-}{n}$,
and hence,
\begin{align}
    Var(\hat{\theta}_{gee}) 
    = \sigma^2(\sum_i X_i^TR^{-1}X_i)^{-1}
    = \frac{\sigma^2}{v^TR^{-1}v}(\sum_ix_i x_i^T)^{-1} \label{eq:GEE_var}.
\end{align}

For snapshot regression, we know $Var(\hat{\theta}_{reg}) = \sigma^2\left ( \sum_i x_ix_i^T\right )^{-1}$, so we have
\begin{align}
    r(\hat{\theta}_{gee}, \hat{\theta}_{reg})=v^TRv \label{eq:GEE_Linear_Eff}
\end{align}

Now we will show $v^TRv>1$. Observe that $v^TR^{-1}v=\langle R^{-0.5}v, R^{-0.5}v\rangle$. Let $a=R^{0.5}v$ and $b=R^{-0.5}v$, we have
\begin{align*}
    |v^Tv|^2\leq (v^TR^{-1}v)( v^TRv)
\end{align*}
by Cauchy-Schwarz inequality, i.e., $|\langle a,b\rangle|^2\leq \langle a, a\rangle\langle b, b\rangle$.
Since $v^Tv=T$ and $v^TRv=\sum_i\sum_j R_{ij}<T^2$, we know:
\begin{align}
    v^TR^{-1}v\geq\frac{T^2}{v^TRv}>1 \label{eq:GEE_inequal}
\end{align}

To further illustrate the connection of efficiency on correlation of repeated measurement, we can assume simple compound symmetric matrix: $R=(1-\rho)I_T + \rho vv^T$. By Woodbury matrix identity, we know $R^{-1} = \frac{1}{1-\rho}(I_T-\frac{\rho}{1+(T-1)\rho}vv^T)$, hence,
\begin{align*}
    v^TR^{-1}v=\frac{1}{1-p}(T-\frac{\rho T^2}{1+(T-1)\rho})=\frac{T}{1+(T-1)\rho}.
\end{align*}
We can see as $\rho\rightarrow1$, $r(\hat{\theta}_{gee}, \hat{\theta}_{reg}) \rightarrow 1$. And as $\rho\rightarrow0$, $r(\hat{\theta}_{gee}, \hat{\theta}_{reg}) \rightarrow T$.

In fact, for general case of $R$, we can define average correlation among different time point as $\bar{\rho} = \frac{1}{T(T-1)}\sum_{i\neq j}R_{ij}$, then from equation~\eqref{eq:GEE_inequal}, we know
\begin{align*}
    r(\hat{\theta}_{gee}, \hat{\theta}_{reg})\geq \frac{T^2}{v^TRv}=\frac{T}{1+(T-1)\bar{\rho}}
\end{align*}

\section{Asymptotics and Efficiency of U test}
\subsection{Pitman Efficiency of MWU over t test}
\subsubsection{Pitman efficiency under specific distributions}
\label{sec:UPitmanExample}
We'll compute pitman efficiency for two heavy tailed distributions. Efficiency computation for other distribution can be found in \cite{blair1980}.

For \textbf{lognormal distribution}: $Log(0,b^2)$, with density $f(y)=\frac{1}{yb\sqrt{2\pi}}\exp(-\frac{(\log y)^2}{2b^2})$ and variance $Var(y)=(e^{b^2}-1)e^{b^2}$, we have
\begin{align*}
    f^2(y)&=\frac{1}{y^2b^22\pi}\exp(-\frac{(\log y)^2}{b^2}), \\
    \int f^2(y)dy &= \int \frac{1}{2\pi b^2}e^{-2u}e^{-u^2/b^2}e^udu\\ 
    &=\frac{1}{2\pi b^2}\int e^{-u-\frac{u^2}{b^2}}du \qquad \text{(let $u=\log y$)}\\
    &= \frac{1}{2\pi b^2}\int e^{\frac{b^2}{4}} e^{-(\frac{u}{b}+\frac{b}{2})^2}du\\
    &=\frac{e^{\frac{b^2}{4}}}{2\pi b^2} \int e^{-w^2}d(bw) \qquad \text{(let $w=\frac{u}{b}+\frac{b}{2}$})\\
    &=\frac{1}{2b\sqrt{\pi}}e^{\frac{b^2}{4}}
\end{align*}
and hence,
\begin{align}
    r(U,\tau)=12(e^{b^2}-1)e^{b^2}(\frac{1}{2b\sqrt{\pi}}e^{\frac{b^2}{4}})^2 = \frac{3}{\pi b^2}(e^{\frac{5}{2}b^2}-e^{\frac{3}{2}b^2}),
\end{align}
which increase exponentially with $b^2$.

For \textbf{Cauchy distribution}: $Cau(0,1)$, with density $f(y)=\frac{1}{\pi(1+y^2)}$, we  have
\begin{align*}
    \int f^2(y)dy &= \frac{1}{\pi^2}\int \frac{1}{(1+y)^2}dy \\
    &= \frac{1}{\pi^2}\int_0^{\frac{\pi}{2}}\cos^2\theta d\theta \qquad \text{(let $y=\cos\theta$)}\\
    &=\frac{1}{\pi^2}\frac{\pi}{2} = \frac{1}{2\pi} \qquad \text{(observing $\cos^2 \theta = \frac{1+\cos(2\theta)}{2}$)}
\end{align*}
and $Var(y)=\infty$, and hence,
\begin{align}
    r(U,\tau) = \infty.
\end{align}

\subsection{Asymptotics of Zero Trimmed U}
\label{sec:ZtuPitAsym}
Let $s_1$ be the sum of ranks of all positive value in the 1st sample, i.e.,
\begin{align*}
    s_1=\sum_i^{n'_1}\kappa(y'_{1i})I_{y'_{1i}>0}.
\end{align*}
Note $\kappa(y'_{1i})=\kappa(y_{1i}),\forall y_{1i}>0$. 

Define,
\begin{align*}
    S'&=\sum_i^{n'_1} \kappa(y'_{1i})\\
    S&=\sum_i^{n_1} \kappa(y_{1i}).
\end{align*}

Observing that $n'_1-n_1^+$ representing number of zeros in $\{y'_{1i}\}_{i=0}^{n'_1}$, and the average rank for those zeros are $\frac{n'_0+n'_1+1+n_0^++n_1^+}{2}$, we have
\begin{align*}
    S'=s_1+(n'_1-n_1^+)\frac{n'_0+n'_1+1+n_0^++n_1^+}{2}.
\end{align*}
Similarly,
\begin{align*}
    S=s_1+(n_1-n_1^+)\frac{n_0+n_1+1+n_0^++n_1^+}{2}.
\end{align*}

And by definition,
\begin{align*}
    W'=S'-\frac{n'_1(n'_0+n'_1+1)}{2},\\
    W=S-\frac{n_1(n_0+n_1+1)}{2}.
\end{align*}
Now, define $w_1=s_1-\frac{n^+_1(n_0^++n_1^++1)}{2}$, we have
\begin{align*}
    W'&=s_1+(n'_1-n_1^+)\frac{n'_0+n'_1+1+n_0^++n_1^+}{2} - \frac{n'_1(n'_0+n'_1+1)}{2} \notag\\
    &= s_1 - \frac{n^+_1(n_0^++n_1^++1)}{2} + \frac{n'_1n_0^+ - n^+_1n'_0}{2} \notag\\
    &=w_1+\frac{n'_1n_0^+ - n^+_1n'_0}{2}
\end{align*}

Similarly, 
\begin{align*}
    W = w_1+\frac{n_1n_0^+ - n^+_1n_0}{2}
\end{align*}
If $d=0$, i.e., $P(y_1^+\geq y_0^+)=\frac{1}{2}$, we have $E(s_1|p_0,p_1)=\frac{n^+_1(n_0^++n_1^++1)}{2}$, i.e., $E(w_1|p_0,p_1)=0$. Then, we have
\begin{align*}
    E(W'|p_0,p_1)&=\frac{n'_1n_0^+ - n^+_1n'_0}{2}=\frac{n_1n_0}{2}(pp_0-pp_1),\\
    E(W|p_0,p_1)&=\frac{n_1n_0^+ - n^+_1n_0}{2}=\frac{n_1n_0}{2}(p_0-p_1).
\end{align*}

Given $p_0$ and $p_1$ are fixed, we know $n_0^+$, $n_1^+$, $n'_0$ and $n'_1$ are all fixed. So,
\begin{align*}
    Var(W|p_0,p_1)=Var(W'|p_0,p_1)=Var(s_1|p_0,p_1)=\frac{n_0^+n_1^+(n_0^+ + n_1^+ + 1)}{12}.
\end{align*}

Then we can compute $Var(W')$ under $H_0$, from its conditional expectation and conditional variance,
\begin{align}
    Var(W') =& Var(E(W'|p_0,p_1)) + E(Var(W'|p_0,p_1)) \notag\\
    =&\frac{n_0^2n_1^2}{4}p^2\left ( \frac{p_1(1-p_1)}{n_1} + \frac{p_0(1-p_0)}{n_0}\right ) \notag\\ 
    & + \frac{n_1n_0p_1p_0}{12}(n_1p_1+n_0p_0)+o(n^3)\\
    =&\frac{n_0n_1(n_0+n_1)}{4}\left [p^3-p^4+\frac{p^3}{3} \right] \notag\\
    &+ o(n^3). \qquad \text{(under $H_0$, $p=p_0=p_1$)} \label{eq:ZTUH0Var}
\end{align}

Similarly, we have
\begin{align}
    Var(W)=&\frac{n_0^2n_1^2}{4}\left ( \frac{p_1(1-p_1)}{n_1} + \frac{p_0(1-p_0)}{n_0}\right ) \notag \\ 
    &+ \frac{n_1n_0p_1p_0}{12}(n_1p_1+n_0p_0)+o(n^3)\\
    =&\frac{n_0n_1(n_0+n_1)}{4}\left [p-p^2+\frac{p^3}{3} \right] + o(n^3). \label{eq:UH0Var}
\end{align}

\subsection{Pitman Efficiency of Zero Trimmed U test over standard U test}
\label{sec:ZtuPitEff}
We have compound alternative hypothesis on two dimension, $m=p_1-p_0$ and $d=P(y_1^+>y_0^+)-\frac{1}{2}$. However, Pitman efficiency is defined for simple hypothesis testing. To handle the compound hypothesis, we specify a direction $\phi$, and on direction of $\phi$, the test would be simple hypothesis. Specifically, let 
\begin{align*}
    m(\nu)&=\nu \cos \phi,\\
    d(\nu)&=\nu \sin \phi.
\end{align*}
On direction of $\phi$, we test 
\[H_0: \nu=0, \text{ vs } H_1^{\phi}:\nu>0.\]

Then we know,
\begin{align}
    \mu'(0)&=\frac{\partial \mu}{\partial \nu}\big |_{\nu=0}=\left ( \frac{\partial \mu}{\partial m} \frac{\partial m}{\partial \nu} + \frac{\partial \mu}{\partial d} \frac{\partial d}{\partial \nu}\right) \big |_{\nu=0} \notag\\
    &= \cos \phi \left(\frac{\partial \mu}{\partial m} \big |_{m=0, d=0}\right) + \sin \phi \left(\frac{\partial \mu}{\partial d} \big |_{m=0, d=0}\right) \label{eq:H1direction}
\end{align}
So, we need to compute $\mu(m,d)$ under local alternative to obtain above quantity. 
Observe that,
\begin{align*}
    w_1=s_1-\frac{n^+_1(n_0^++n_1^++1)}{2} = n_0n_1(\frac{1}{2}-U_{n_0^+n_1^+})
\end{align*}
where, $U_{n_0^+n_1^+}$ is Mann-Whitney U on positive-only samples: 
\begin{align*}
    U_{n_0^+n_1^+} = \frac{1}{n_0^+n_1^+}\sum_i^{n_1^+} \sum_j^{n_0^+} I_{y'_{1i}>y'_{0j}}
\end{align*}
Knowing that $E(U_{n_0^+n_1^+}|p_0,p_1)=P(y^+_1\geq y_0^+)$, we have
\begin{align*}
    E(W'|p_0,p_1)&=-n_0^+n_1^+\left [P(y^+_1\geq y_0^+) - \frac{1}{2}\right] + \frac{n'_1n_0^+ - n^+_1n'_0}{2}\\
    &= -n^+_0n^+_1d - \frac{n^+_1 n'_0 - n'_1 n^+_0}{2}
\end{align*}
Hence, 
\begin{align*}
    \mu_{W'}(m,d)&=E(W')=E(E(W'|p_0,p_1))\\
    &=-n_1n_0dp(p+m)-\frac{n_1n_0}{2}\left[ (p+m)^2-p(p+m)\right]\\
    &=-\frac{n_1n_0}{2} \left[ 2p(p+m)d + m(p+m)\right].
\end{align*}

Similarly,
\begin{align*}
    \mu_{W}(m,d)&=E(W)=E(E(W|p_0,p_1))\\
    &=E \left( -n^+_0n^+_1d - \frac{n^+_1 n_0 - n_1 n^+_0}{2} \right)\\
    &=-n_1n_0dp(p+m)-\frac{n_1n_0}{2}\left[ p+m-p\right]\\
    &=-\frac{n_1n_0}{2} \left[ 2p(p+m)d + m\right].
\end{align*}

We can ignore term $-\frac{n_0n_1}{2}$ for the ratio $\frac{\mu'_{W'}(0)}{\mu'_{W}(0)}$. Observe that
\begin{align*}
    \frac{\partial \mu_{W'}}{\partial m}\big |_0 &= \left( 2pd+p+2m \right)\big|_{m=0,d=0} =p,\\
    \frac{\partial \mu_{W'}}{\partial d}\big |_0 &= \left( 2p(p+m) \right)\big|_{m=0,d=0} =2p^2,\\
    \frac{\partial \mu_{W}}{\partial m}\big |_0 &= \left( 2pd+1 \right)\big|_{m=0,d=0} =1,\\
    \frac{\partial \mu_{W'}}{\partial d}\big |_0 &= \left( 2p(p+m) \right)\big|_{m=0,d=0} =2p^2.\\
\end{align*}
Combining above with~\eqref{eq:ZTUH0Var},~\eqref{eq:UH0Var} and~\eqref{eq:H1direction}, we complete the proof of the pitman efficiency for Zero Trimmed U:
\begin{align*}
    r^{\phi}(W',W) = \frac{\sigma_W^2(0)}{\sigma_{W'}^2(0)}\left (\frac{\mu'_{W'}(0)}{\mu'_W(0)} \right )^2
    =\frac{1-p+\frac{p^2}{3}}{p^2-p^3+\frac{p^2}{3}}\left ( \frac{p\cos \phi+2p^2\sin \phi}{\cos \phi + 2p^2 \sin \phi} \right )^2.
\end{align*}

Note that we actually used the adjusted variance for non-zero trimmed version $W$ to handles the ties on the zeros. If we calculated the unadjusted variance from the original approach, i.e., $Var(W^{o})=\frac{n_1n_2(n_1+n_2+1)}{12}$, then we have pitman efficiency for Zero-Trimmed U over unadjusted W as:

\begin{align}
    r^{\phi}(W',W^o) 
    =\frac{\frac{1}{3}}{p^3-p^4+\frac{p^3}{3}}\left ( \frac{p\cos \phi+2p^2\sin \phi}{\cos \phi + 2p^2 \sin \phi} \right )^2, \label{eq:r_ww_o}
\end{align}
observing that $W=W^o$ for point estimate.

\section{Doubly Robust Generalized U}
\label{apx:DRU}
\subsection{The Robustness of DRGU}
\label{apx:DRU_Unbias}
When there are no confounding effects, i.e., $y\perp z$, 
we can show that $E(h(y_i, y_j))=\delta$ by conditioning on $z$:
\begin{align*}
    E(h(y_i, y_j)) &= P(z_i=1)E(h_{ij}|z_i=1) + P(z_i=0)E(h_{ij}|z_i=0) \\
    &= p(\frac{1-p}{2p(1-p)}\delta+0) + (1-p)(0+\frac{p}{2p(1-p)}\delta) \\
    & = \delta,
\end{align*}
and hence $E(U_n)=\delta$. We can further show asymtotic normality: $\sqrt{n}(U_n - \delta) \rightarrow_d N(0, 4\sigma^2_h)$.

When there are confounding effects, we can form a inverse probability weighted (IPW) U statistics:

\[U_n^{IPW} = \left[ \binom{n}{2}\right]^{-1} \sum_{i,j\in C^n_2 } h^{IPW}_{ij},\]

where,
\[h_{ij}^{IPW} = \frac{z_i(1-z_j)}{2\pi_i(1-\pi_j)}\varphi(y_{i1}- y_{j0}) + \frac{z_j(1-z_i)}{2\pi_j(1-\pi_i)}\varphi(y_{j1} - y_{i0}),\]
and $\pi_i = E(z_i|w_i).$ 

Assuming $y \perp z | w$, we can show, 
\begin{align*}
    E(h_{ij}^{IPW}) =& E(E(h_{ij}^{IPW}|w_i, w_j))\\
    =& E(\frac{E(z_i(1-z_j)\varphi(y_{i1} - y_{j0}) | w_i, w_j)}{2\pi_i(1-\pi_j)}) \\& +E(\frac{E(z_j(1-z_i)\varphi(y_{j1} - y_{i0}) | w_i, w_j)}{2\pi_j(1-\pi_i)}) \\
    =& E(\frac{E(z_i(1-z_j)|w_i, w_j)E(\varphi(y_{i1} - y_{j0}) | w_i, w_j)}{2\pi_i(1-\pi_j)}) \\
    &+E(\frac{E(z_j(1-z_i)|w_i, w_j)E(\varphi(y_{j1} - y_{i0}) | w_i, w_j)}{2\pi_j(1-\pi_i)}) \\
    =& \frac{\pi_i(1-\pi_j)}{2\pi_i(1-\pi_j)}E(\varphi(y_{i1} - y_{j0})) + \frac{\pi_j(1-\pi_i)}{2\pi_j(1-\pi_i)}E(\varphi(y_{j1} - y_{i0}))\\
    =& \delta,
\end{align*}
and hence the IPW adjusted U statistics is unbiased, i.e., $E(U_n^{IPW})=\delta$.

By further introducing $g_{ij}=E(\varphi(y_{i1} - y_{j0})|w_i, w_j)$, we form a Doubly Robust Generalized U statistics, $U_n^{DR}$, with kernel,
\begin{align*}
    h_{ij}^{DR} =& \frac{z_i(1-z_j)}{2\pi_i(1-\pi_j)}(\varphi(y_{i1} - y_{j0})-g_{ij}) \\ 
    &+ \frac{z_j(1-z_i)}{2\pi_j(1-\pi_i)}(\varphi(y_{j1} - y_{i0})-g_{ji})+\frac{g_{ij}+g_{ji}}{2}.
\end{align*}
It's easy to show that $E(h_{ij}^{DR})=\delta$, observing
\begin{align*}
    E(h_{ij}^{DR}) =& E(E(h_{ij}^{DR}|w_i, w_j))\\
    =& E(\frac{E(z_i(1-z_j)|w_i, w_j)(g_{ij}-g_{ij})}{2\pi_i(1-\pi_j)}) \\
    &+ E(\frac{E(z_j(1-z_i)|w_i, w_j)(g_{ji} - g_{ji})}{2\pi_j(1-\pi_i)}) + E(\frac{g_{ij}+g_{ji}}{2})\\
    =& 0+0+\delta =\delta.
\end{align*}

\subsection{Semi-parametric Efficiency of DRGU}
\label{apx:DRU_Efficiency}
In this section, we sketch the proof for DRGU as most efficient estimator under semi-parametric set-up. 

At a high level, we need to show DRGU has influence function (IF) that correspond to efficient influence function (EIF) for parameter $\delta = \varphi(y_1-y_0)$, so naturally there are two steps: \\
(i) find EIF for $\delta=\varphi(y_1-y_0)$, \\
(ii) show DRU's IF is consistent with EIF.

\textbf{\textit{Preliminary}}: For regular asymptotic linear estimator $\hat{\theta}$, we have $\sqrt{n}(\hat{\theta}-\theta)=\frac{1}{n}\sum_i \vartheta_i + o_p(1)$. $\vartheta$ is the IF for $\hat{\theta}$. EIF $\vartheta'$ is defined as the unique IF with smallest variance, i.e., $Var(\vartheta')\leq Var(\vartheta), \forall \vartheta$. Since  $\sqrt{n}(\hat{\theta}-\theta) \rightarrow_p N(0, Var(\vartheta)$), we know estimator with EIF has smallest variance. 

For finding the EIF, we follow the standard recipe in semi-parametric theory (i.e., 13.5 of \cite{tsiatis2006semiparametric}).
\begin{enumerate}
    \item Identify IF $\vartheta^F$ for full data, $O^F=\{(y(1), y(0), x)\}$, where $y(1)$ and $y(0)$ represent response variable under treatment and control respectively.
    \item Find all IFs $\vartheta$ for observation data $O^o=\{(y,z,w)\}$,
    \begin{align*}
        \vartheta(y,z,x) = \vartheta^o(y,z,x) + \Lambda
    \end{align*}
    where $E(\vartheta^o(y,z,x)|O^F)=\vartheta^F(y_(1),y(0))$ and $\Lambda=\{L: E(L(y,z,x)|O^F)=0\}$ is the augmentation space.
    Note that here, $y=zy(1)+(1-z)y(0)$ with the stable unit treatment value assumption(SUTVA).
    \item Identify the EIF through projection onto the augmentation space.
    \begin{align*}
        \vartheta'(y,z,x) = \vartheta^o(y,z,x) - \Pi(\vartheta^o(y,z,x)|\Lambda)
    \end{align*}
    where $\Pi(f|\Lambda)$ is a projection of a function $f$ on space $\Lambda$, such that $E[(f-\Pi(f|\Lambda))g]=0, \forall g\in \Lambda$.
\end{enumerate}
For full data $O^F=\{(y(1), y(0),x)\}$, we can construct U kernel 
\[h^F_{ij} = 0.5(\varphi(y_i(1)-y_j(0)+\varphi(y_j(1)-y_i(0)),\] 
and form a U statistic: 
\[U^F=\binom{n}{2}^{-1}\sum_{i\neq j} h_{ij}^F\]
for unbiased estimation of $\delta=\varphi(y(1)-y(0))$.

From Hajek projection of U statistics, we know $\sqrt{n}(U^F -\delta)=\frac{2}{n}\sum_i \tilde{h}(y_i) + o_p(1)$, where $\tilde{h}(y_i) = E(h^F_{ij}|O_i^F)-\delta$. 

Now observe, 
\begin{align*}
    E(h^F_{ij}|O_i^F) &= 0.5E(\varphi(y_i(1)-y_j(0)|O^F_i) + 0.5E(\varphi(y_j(1)-y_i(0)|O^F_i)\\
    &=0.5\int \varphi(y_i(1) - s)p_0(s)ds + 0.5\int \varphi(t - y_i(0))p_1(t)dt\\
    &= 0.5 h_1(y_i(1)) + 0.5h_0(y_i(0))
\end{align*}
where $h_1(y) = \int \varphi(y-s)p_0(s)ds$, $h_0(y) = \int \varphi(t-y)p_1(t)dt$, and $p_1(\cdot), p_0(\cdot)$ are marginal density of $y$ under treatment and control respectively.

We then have $\sqrt{n}(U^F -\delta)=\frac{1}{n}\sum_i [ h_1(y_i(1)) + h_0(y_i(0)) - 2\delta] + o_p(1)$, and as a result the corresponding IF under full data is $\vartheta^F = h_1+h_0-2\delta$.

Next step is to find an IF $\vartheta^o$ for observation data $O^o= \{(y,z,x)\}$. Let $\vartheta^o$ be the inverse propensity weighting version of the $\vartheta^F$, i.e., 
\[\vartheta^o = \frac{z}{\pi} h_1 + \frac{1-z}{1-\pi}h_0 - 2\delta\]
where $\pi=E(z|x)$.

We can verify that $E(\vartheta^o|O^F)=\vartheta^F$, observing
\begin{align*}
    E(\frac{z}{\pi}h_1|O^F) = \frac{h_1}{\pi}E(z|x) =h_1
\end{align*}
as similarly $E(\frac{1-z}{1-\pi}h_0|O^F)=h_0$.

We then specify the augmentation space $\Lambda$. For any function $L(y,z,x)$, since $z\in\{0,1\}$, we can represent the function as $L(y,z,w)=zL_1(y,w)+(1-z)L_0(y,w)$. Further by definition, $E(L|O^F)=0$, we know
\begin{align*}
    E(L|O^F) = \pi L_1(y(1), w)+(1-\pi)L_0(y(0),w)=0, \forall w,y(0),y(1)
\end{align*}

Since above equation applies to all values of $w,y(0),y(1)$, we know $L_0(y(0),w)=L_0(w)$, $L_1(y(1),w)=L_1(w)$, $L_0(w)=\frac{-\pi}{1-\pi}L_1(w)$, and we can represent $L(y,z,w)$ as
\[L(y,z,w) = zL_1(w) + (1-z)\frac{-\pi}{1-\pi}L_1(w)=\frac{z-\pi}{1-\pi}L_1(w)\]
Thus, we can specify $\Lambda = \{L: L(y,z,w)=(z-\pi)f(w) \text{ for arbitrary } f\}$.

We next find projection so that EIF $\vartheta' = \vartheta^o - \Pi(\vartheta^o|\Lambda)$. From specification of $\Lambda$, let $\Pi(zh_1|\Lambda)=(z-\pi)f_1$, and $\Pi((1-z)h_0|\Lambda)=(z-\pi)f_0$.
By definition,
\[E([zh_1 -(z-\pi)f_1][(z-\pi)f])=0, \forall f.\]

Observing,
\begin{align*}
    &E([zh_1 -(z-\pi)f_1][(z-\pi)f])\\ 
    =& E(z(z-\pi)fh - (z-\pi)^2f_1f)\\
    =&E(\pi(1-\pi)fE(h_1|z=1,x)) - E(\pi(1-\pi)f_1f) \\
    =&E(\pi(1-\pi)\left[ E(h_1|z=1,x) -f_1 \right]f) =0, \forall f
\end{align*}
we have $f_1 = E(h_1|z=1,x)$.
Similarly, we have $f_0 = -E(h_0|z=0,x)$. Substitute the two equation, we get 
\begin{align*}
    \Pi(\vartheta^o|\Lambda) = \frac{z-\pi}{\pi}E(h_1|z=1,x) - \frac{z-\pi}{1-\pi}E(h_0|z=0,x)
\end{align*}
and hence the EIF is
\begin{align}
    \vartheta' =& \frac{z}{\pi} h_1 + \frac{1-z}{1-\pi}h_0 - 2\delta - \frac{z-\pi}{\pi}E(h_1|z=1,x) + \frac{z-\pi}{1-\pi}E(h_0|z=0,x) \notag\\
    = &\frac{z}{\pi}(h_1 - E(h_1|z=1,w)) + \frac{1-z}{1-\pi}(h_0-E(h_0|z=0,w)) \notag\\
    &+ E(h_1|z=1,w) + E(h_0|z=0,w) - 2\delta
\end{align}

We then need to show the $U_n^{DR}$ has influence function that is consistent with $\vartheta'$, i.e., $\vartheta^{DR}=\vartheta + o_p(1)$.
From Hajek projection, we can obtain $U_n^{DR}$'s influence function, i.e., $\vartheta^{DR} = 2E(h_{ij}^{DR}|O^o_i)-2\delta$. 

Recall 
\begin{align*}
    h_{ij}^{DR} =& \frac{z_i(1-z_j)}{2\pi_i(1-\pi_j)}(\varphi(y_{i1} - y_{j0})-g_{ij}) \\
    &+ \frac{z_j(1-z_i)}{2\pi_j(1-\pi_i)}(\varphi(y_{j1} - y_{i0})-g_{ji})+\frac{g_{ij}+g_{ji}}{2}.
\end{align*}

Let's calculate the $E(h_{ij}^{DR}|O^o_i)$ term by term.

For the first term, we have
\begin{align*}
    E(z_i\frac{1-z_j}{1-\pi_j}\varphi(y_{i1}-y_{j0}))|O^o_i) = E((1-z_i)\varphi(y_{i1}-y_{j0}))|O^o_i) = z_ih_1(y_{i})
\end{align*}
and similarly $E((1-z_i)\frac{z_j}{\pi_j}\varphi(y_{j1}-y_{i0}) = (1-z_i)h_0(y_{i})$

By definition, $g_{ij}=E[\varphi(y_i-y_j)|w_i,w_j,z_i=1,z_j=0]$ and $g_{ji}=E[\varphi(y_j-y_i)|w_j,w_i,z_j=1,z_i=0]$. we can show

\begin{align*}
    E(g_{ij}|O^o_i) &= \left . \int \left [ \int \int \varphi(s-t)p_1(s|w_i)p_0(t|w_j) dsdt  \right ]p(w_j)dw_j \right|_{s=y_i}\\
    &= \left .\int \int \varphi(s-t)p_1(s|w_i) \left [\int  p_0(t|w_j)p(w_j)dw_j\right ]dsdt \right|_{s=y_i} \\
    &= \left .\int \int \varphi(s-t)p_1(s|w_i) p_0(t)dsdt \right|_{s=y_i}\\
    &= \left .\int \left [ \int \varphi(s-t) p_0(t)dt \right] p(s|w_i,z_i=1)ds \right|_{s=y_i}\\
    & = E(h_1(y_i)|w_i,z_i=1)
\end{align*}
and similarly, $E(g_{ji}|O^o_i)=E(h_0(y_i)|w_i,z_i=0)$.

We also know 
\begin{align*}
    E(\frac{1-z_j}{1-\pi_j}g_{ij}|O^o_i) = E\left [ E(\frac{1-z_j}{1-\pi_j}|w_j)g_{ij}|O^o_i\right] = E(g_{ij}|O^o_i).
\end{align*}
and similarly $E(\frac{z_j}{\pi_j}g_{ji}|O^o_i) = E(g_{ji}|O^o_i)$.

Substituting above equations, we have 
\[E(h_{ij}^{DR}|O^o_i) = \frac{\vartheta'}{2} + \delta,\] and hence $\vartheta^{DR} = \vartheta'$ exactly.

\subsection{Asymptotics of DRGU with UGEE}
\label{apx:DRU_Asymp}

We'll first sketch the proof for the \textbf{\textit{asymptotic normality}} of DRGU.

Recall that, 
\[\mathbf{U}_n(\theta) = \sum_{i,j\in C^n_2} \mathbf{U}_{n,ij} = \sum_{i,j\in C^n_2} \mathbf{G}_{ij}(\mathbf{h}_{ij}-\mathbf{f}_{ij}) = \mathbf{0},\]
and 
\begin{align*}
    \mathbf{G}_{ij} &=  \mathbf{D}_{ij}^T \mathbf{V}^{-1}_{ij} \\
    \mathbf{D}_{ij} &= \frac{\partial \mathbf{f}_{ij}}{\partial \theta}, \\
    \mathbf{V}_{ij} &= AR(\alpha)A.
\end{align*}

Recall $\mathbf{u}_{i} = E(\mathbf{U}_{n,ij}|y_{i0},y_{i1},z_i, w_i)$, $\Sigma=Var(\mathbf{u}_i)$, $\mathbf{M}_{ij} = \frac{\partial (\mathbf{f}_{ij}-\mathbf{h}_{ij})} {\partial \theta}$, and $\mathbf{B} = E(\mathbf{G}\mathbf{M})$, and $\hat{\delta}$ be the 1st element in $\hat{\theta}$.

Let $\bar{U}_n(\theta, \alpha) = \frac{1}{\binom{n}{2}}\sum_{i,j\in C^n_2} U_{n,ij}$. We know $\bar{U}_n(\theta, \alpha)$ is a U statistics with mean $E(U_{n,ij}) = 0$. From asymptotic theory of U statistics, we know
\begin{align*}
    \sqrt{n}\bar{U}_n(\theta, \alpha) \rightarrow_d N(0,4\Sigma)
\end{align*}

Let $\hat{\alpha}$ be the estimate of $\alpha$ for the working correlation $R(a)$, and assume mild regularity condition: $\sqrt{n}(\hat{\alpha}-\alpha)=O_p(1)$. And let $\hat{\theta}$ be the estimate of the $\theta$, i.e., $\bar{U}_n(\hat{\theta}, \hat{\alpha})=0$. Observing the following Taylor expansion, 
\begin{align*}
    0 &= \bar{U}_n(\hat{\theta}, \hat{\alpha})\\
    &= \bar{U}_n(\theta, \alpha) + \frac{\partial \bar{U}_n(\theta,\alpha)}{\partial \theta}(\hat{\theta}-\theta) + \frac{\partial \bar{U}_n(\theta, \alpha)}{\partial \alpha}(\hat{\alpha} - \alpha) + o_p(n^{-0.5}),
\end{align*}

we know,
\begin{align}
    \sqrt{n}\bar{U}_n(\theta, \alpha) = -\sqrt{n}\frac{\partial \bar{U}_n}{\partial \theta} (\hat{\theta}-\theta) - \sqrt{n}\frac{\partial \bar{U}_n}{\partial \alpha}(\hat{\alpha}-\alpha) + o_p(1). \label{eq:GEETaylor}
\end{align}
Since $E(\mathbf{h}_{ij}-\mathbf{f}_{ij})=0$, we know $E(\frac{\partial U_{n,ij}}{\partial\alpha}) = 0$, and hence $\frac{\partial \bar{U}_n}{\partial \alpha} = o_p(1)$. Combining with the regularity condition $\sqrt{n}(\hat{\alpha}-\alpha)=O_p(1)$, we have
\begin{align}
    \sqrt{n}\frac{\partial \bar{U}_n}{\partial \alpha}(\hat{\alpha}-\alpha) = o_p(1)O_p(1) = o_p(1).
\end{align}
Then equation~\eqref{eq:GEETaylor} reduce to,
\begin{align}
    \sqrt{n}(\hat{\theta }- \theta) = -(\frac{\partial \bar{U}_n}{\partial \theta})^{-}\sqrt{n}\bar{U}_n(\theta, \alpha) + o_p(1) \label{eq:UGEE_inv}
\end{align}
where $(\cdot)^-$ denote general inverse.


Let $S_{ij} =  \mathbf{h}_{ij} - \mathbf{f}_{ij}$. Given that $\frac{\partial \bar{U}_n}{\partial \theta}=\frac{1}{\binom{n}{2}}\sum_{i>j} \frac{\partial G_{i,j}S_{i,j}}{\partial \theta}$,  we have
\begin{align}
    \frac{\partial \bar{U}_n}{\partial \theta} \rightarrow_p E(G\frac{\partial S}{\partial \theta})= -E(GM) \label{eq:UGEE_B}
\end{align}

To obtain equation~\eqref{eq:UGEE_B}, we observe that 
\begin{align*}
    \frac{\partial \bar{U}_n}{\partial \theta} &=\frac{1}{\binom{n}{2}}\sum_{i>j} \frac{\partial G_{ij}S_{ij}}{\partial \theta} \\
    &= \frac{1}{\binom{n}{2}}\sum_{i>j} \frac{\partial D_{ij}^T}{\partial \theta} V_{ij}^{-1}S_{ij} + \frac{1}{\binom{n}{2}}\sum_{i>j} 
 D_{ij}^TV_{ij}^{-1}\frac{\partial S_{ij}}{\partial \theta}.
\end{align*}
Since $\frac{1}{\binom{n}{2}}\sum_{i>j} S_{ij}\rightarrow_p 0$, we have negligible first term. As a result, 
\begin{align*}
    \frac{\partial \bar{U}_n}{\partial \theta} &= o_p(1) + \frac{1}{\binom{n}{2}}\sum_{i>j} 
 D_{ij}^TV_{ij}^{-1}\frac{\partial S_{ij}}{\partial \theta} \rightarrow_p -E(GM)
\end{align*}

Combining equation~\eqref{eq:UGEE_inv} and equation~\eqref{eq:UGEE_B}, we have 
\begin{align}
    \sqrt{n}(\hat{\theta}-\theta) = B^-\sqrt{n}\bar{U}_n(\theta, \alpha) + o_p(1)
\end{align}
where $B=E(GM)$. Hence, we establish the following asymptotic normality:
\begin{align*}
    \sqrt{n}(\hat{\theta} - \theta)\rightarrow_d N(0, 4B^{-}\Sigma B^{-T}).
\end{align*}

We skip the proof for \textit{\textbf{consistency}} when only one of $\pi$ and $g$ is correctly specified, as most of it has been discussed in Appendix ~\ref{apx:DRU_Unbias}. 

As for \textit{\textbf{semi-parametric bound}} of $\hat{\delta}$, proof is straightforward building on results from ~\ref{apx:DRU_Efficiency}.

Observing $D$ has structure of block diagonal with following structure:
\begin{align*}
    D = \begin{bmatrix}
            1 & 0 & \cdots & 0 \\
            0 & d_{22} & \cdots & d_{2p} \\
            \vdots & \vdots & \ddots & \vdots \\
            0 & d_{T2} & \cdots & d_{Tp}
        \end{bmatrix}
\end{align*}

Recall EIF $\vartheta'$ from ~\ref{apx:DRU_Efficiency}, we know $E(\vartheta'S_{\pi})=0$ and $E(\vartheta'S_{g})=0$, and thus $E(M)$ has following structure:
\begin{align*}
    E(M) = \begin{bmatrix}
            1 & 0 & \cdots & 0 \\
            0 & m_{22} & \cdots & m_{2p} \\
            \vdots & \vdots & \ddots & \vdots \\
            0 & m_{T2} & \cdots & m_{Tp}
        \end{bmatrix}
\end{align*}

We then know $B=E(GM)$ has the following structure:

\begin{align*}
    B = \begin{bmatrix}
            \sigma_1^{-2} & 0 & \cdots & 0 \\
            0 & b_{22} & \cdots & b_{2p} \\
            \vdots & \vdots & \ddots & \vdots \\
            0 & b_{p2} & \cdots & b_{pp}
        \end{bmatrix}
\end{align*}

Since $E(\vartheta'S_{\pi})=0$ and $E(\vartheta'S_{g})=0$, we know $\Sigma$ is block diagonal,

\begin{align*}
    \Sigma = \begin{bmatrix}
            \sigma_1^{2} & 0 & \cdots & 0 \\
            0 & s_{22} & \cdots & s_{2p} \\
            \vdots & \vdots & \ddots & \vdots \\
            0 & s_{p2} & \cdots & s_{pp}
        \end{bmatrix}
\end{align*}

Observing the asymptotic covariance matrix is $4 B^{-}\Sigma B^{-T}$, we know asymptotic variance of $\hat{\delta}$ is same as that of EIF, i.e., $\sigma^2_{\delta} = 4\sigma_1^2=Var(\vartheta')$. 

\section{Details on Simulation Studies}
\subsection{Regression Adjustment}
\label{apx:RA_Simu}

We compare the type I error rate of regression adjustment and the unadjusted $t$-test. We perform these simulations using data generated from a Poisson distribution with the following generation process: $ w_i \sim \mathcal{N}(0,1), \: z_i|w_i \sim \textrm{Bernoulli}\left( \frac{1}{1+e^{-\gamma w_i}} \right), \: y_i|z_i,w_i \sim \textrm{Poisson}\left( e^{2 + \beta_z z_i + \beta_w w_i}\right).$

Here $\gamma  \geq 0$ is a hyperparameter which controls the degree of confounding. When $\gamma = 0$ there is no confounding. $\beta_z$ controls the treatment effect. We evaluation type I error with $\beta_z=0$, and power with $\beta_z>0$.

\begin{table}[h]
\centering
\caption{Type I Error Comparison: t-test vs. Regression Adjustment}
\label{tab:type1error}
\begin{tabular}{ccc}
\hline
$\gamma$ & t-test & RA \\
\hline
0.0 & 0.0504 & 0.0504 \\
0.2 & 0.0576 & 0.0482 \\
0.4 & 0.0706 & 0.0522 \\
0.6 & 0.0844 & 0.0496 \\
0.8 & 0.1082 & 0.0570 \\
1.0 & 0.1262 & 0.0524 \\
\hline
\end{tabular}
\end{table}

We validate that regression adjustment controls type I error, and the unadjusted $t$-test leads to type I error rate inflation under confounding. 

\begin{table}[h]
\centering
\caption{Power Comparison: t-test vs. Regression Adjustment}
\label{tab:power_comparison}
\begin{tabular}{c|cc|cc}
\hline
\multirow{2}{*}{Treatment Effect} & \multicolumn{2}{c|}{$\gamma = 0.0$} & \multicolumn{2}{c}{$\gamma = 0.1$} \\
\cline{2-5}
 & t & RA & t & RA \\
\hline
0.10 & 0.727 & 0.702 & 0.819 & 0.690 \\
0.11 & 0.722 & 0.779 & 0.825 & 0.767 \\
0.12 & 0.734 & 0.848 & 0.826 & 0.834 \\
0.13 & 0.729 & 0.907 & 0.828 & 0.879 \\
0.14 & 0.751 & 0.947 & 0.862 & 0.949 \\
0.15 & 0.773 & 0.956 & 0.863 & 0.961 \\
0.16 & 0.795 & 0.975 & 0.881 & 0.987 \\
0.17 & 0.793 & 0.992 & 0.885 & 0.991 \\
0.18 & 0.777 & 0.995 & 0.881 & 0.990 \\
0.19 & 0.796 & 0.999 & 0.900 & 0.996 \\
0.20 & 0.807 & 0.997 & 0.908 & 0.999 \\
\hline
\end{tabular}
\end{table}

We demonstrate that regression adjustment improves power over t-test ($\gamma =0$). When there is confounding present, the power of raw unadjusted $t$-test is not valid as it can not control type I error. 

\subsection{GEE}
\label{apx:GEE_Simu}

We evaluate the Type I error and power of two estimators in the presence of confounding under varying sample sizes and effect sizes.

(i) GLM adjustment at final time point: at $t = T$, fit a Poisson regression  
\[
Y_{iT} \sim \mathrm{Poisson}\!\bigl(\exp(\beta_0 + \beta_1 z_i + \gamma w_i)\bigr)
\quad\Longrightarrow\quad
\hat\beta_1^{\rm GLM}\,. 
\]

(ii) GEE adjustment with longitudinal data: using all observations $t=1,\dots,T$, obtain $\hat\beta_1^{\rm GEE}$ by solving the estimating equation  
\[
U_n(\beta_1)
=\sum_{i=1}^N\sum_{t=1}^T D_{it}^\top\bigl[Y_{it}-\exp(\beta_0 + \beta_1 z_i + \gamma w_i)\bigr]
=0,
\]  
where  
\[
D_{it}
=\frac{\partial\,E[Y_{it}\mid z_i,w_i]}{\partial\beta_1}
=\;z_i\,\exp(\beta_0 + \beta_1 z_i + \gamma w_i)\,.
\]

We generate a longitudinal panel of $N$ subjects over $T$ visits by first drawing a time-invariant confounder and treatment for each subject, then simulating a Poisson count at each visit:
\[
w_i \sim \mathcal{N}(0,1), 
\quad 
z_i \sim \mathrm{Bernoulli}\bigl(\sigma(\alpha_0+\alpha_1 w_i)\bigr), 
\]
\[
Y_{it} \sim \mathrm{Poisson}\!\bigl(\exp(\beta_0 + \beta_1 z_i + \gamma w_i)\bigr),
\]
for $i=1,\dots,N$ and $t=1,\dots,T$.

\begin{table}[h]
\centering
\caption{Empirical Type I error rates ($\beta_1=0$) for GEE and GLM estimators under confounded assignment at nominal levels 
$\alpha$.}
\label{tab:type1_error}
\begin{tabular}{cccc}
\hline
Sample Size & $\alpha$ & GEE & GLM \\
\hline
50  & 0.05 & 0.068 & 0.057 \\
50  & 0.01 & 0.021 & 0.013 \\
200 & 0.05 & 0.048 & 0.044 \\
200 & 0.01 & 0.009 & 0.005 \\
\hline
\end{tabular}
\end{table}
We demonstrate that GEE controls type I error adequately in large samples, with only modest inflation when sample sizes are small.

\begin{table}[h]
\centering
\caption{Empirical power for Poisson GEE vs.\  GLM estimators across sample sizes $N$, effect sizes $\beta_1$, and significance levels $\alpha$.}
\label{tab:power_analysis}
\begin{tabular}{ccccc}
\hline
Sample Size & $\beta_1$ & $\alpha$ & GEE & GLM  \\
\hline
50  & 0.10 & 0.05 & 0.283 & 0.100 \\
50  & 0.10 & 0.01 & 0.138 & 0.025 \\
50  & 0.20 & 0.05 & 0.749 & 0.200 \\
50  & 0.20 & 0.01 & 0.556 & 0.066 \\
200 & 0.10 & 0.05 & 0.729 & 0.189 \\
200 & 0.10 & 0.01 & 0.490 & 0.065 \\
200 & 0.20 & 0.05 & 1.000 & 0.645 \\
200 & 0.20 & 0.01 & 0.997 & 0.404 \\
\hline
\end{tabular}

\end{table}

We demonstrate that by leveraging longitudinal repeated measurements, the GEE‐adjusted estimator achieves higher statistical power than that of Poisson GLM across both small and large samples. Moreover, this power advantage is especially pronounced at medium effect sizes ($\beta_1=0.1$) compared to larger ones ($\beta_1=0.2$).

\subsection{Mann Whitney U}
\label{apx:MWU_Simu}
We compare the zero-trimmed Mann-Whitney U-test to the standard Mann-Whitney U-test and two-sample $t$-test in type I error rate and power. We simulate the three tests using data generated from zero-inflated log-normal and positive Cauchy distributions and multiple effect sizes. Formally, we generate control data $y_{0i} =(1-D_i)y_{0i}'$, where $D_i \sim \textrm{Bernoulli}(p_0)$ and $y_{0i}' \sim f(0, \sigma)$. We generate test data $y_{1j} =(1-D_j)y_{1j}'$  where $D_j \sim \textrm{Bernoulli}(p_0 +p_{\Delta})$ and $y_{1j}' \sim f(\mu, \sigma)$ for $p_{\Delta}, \mu \geq 0$. Here $f$ denotes either the lognormal or positive Cauchy distribution. 

\begin{table}[H]
\centering
\caption{Type I Error Rates at $\alpha=0.05$ for Zero-Inflated Data}
\label{tab:type1error_alpha05}
\begin{tabular}{llc|ccc}
\hline
\multirow{2}{*}{Distribution} & \multirow{2}{*}{$p_0$} & \multirow{2}{*}{$n$} & \multicolumn{3}{c}{Type I Error Rate} \\
 &  &  & ZTU & MWU & t-test \\
\hline
\multirow{8}{*}{LogNormal} & \multirow{2}{*}{0.0} & 50 & 0.0540 & 0.0540 & 0.0015 \\
 &  & 200 & 0.0515 & 0.0515 & 0.0040 \\
\cmidrule{2-6}
 & \multirow{2}{*}{0.2} & 50 & 0.0435 & 0.0500 & 0.0025 \\
 &  & 200 & 0.0480 & 0.0545 & 0.0050 \\
\cmidrule{2-6}
 & \multirow{2}{*}{0.5} & 50 & 0.0315 & 0.0465 & 0.0020 \\
 &  & 200 & 0.0405 & 0.0490 & 0.0055 \\
\cmidrule{2-6}
 & \multirow{2}{*}{0.8} & 50 & 0.0230 & 0.0475 & 0.0005 \\
 &  & 200 & 0.0305 & 0.0455 & 0.0050 \\
\hline
\multirow{8}{*}{Positive Cauchy} & \multirow{2}{*}{0.0} & 50 & 0.0535 & 0.0535 & 0.0220 \\
 &  & 200 & 0.0530 & 0.0525 & 0.0200 \\
\cmidrule{2-6}
 & \multirow{2}{*}{0.2} & 50 & 0.0465 & 0.0540 & 0.0205 \\
 &  & 200 & 0.0405 & 0.0480 & 0.0240 \\
\cmidrule{2-6}
 & \multirow{2}{*}{0.5} & 50 & 0.0335 & 0.0455 & 0.0215 \\
 &  & 200 & 0.0420 & 0.0500 & 0.0175 \\
\cmidrule{2-6}
 & \multirow{2}{*}{0.8} & 50 & 0.0290 & 0.0550 & 0.0230 \\
 &  & 200 & 0.0355 & 0.0500 & 0.0170 \\
\hline
\end{tabular}
\end{table}

\begin{table}[H]
\centering
\caption{Power Comparison for Positive Cauchy and LogNormal Distributions with Equal Zero-Inflation (50\%)}
\label{tab:power_comparison}
\begin{tabular}{ccc|ccc}
\hline
\multirow{2}{*}{Distribution} & \multirow{2}{*}{$n$} & \multirow{2}{*}{Effect Size} & \multicolumn{3}{c}{Power at $\alpha = 0.05$} \\
 & & & ZTU & MWU & t-test \\
\hline
\multirow{8}{*}{Positive Cauchy} & \multirow{4}{*}{50} & 0.25 & 0.038 & 0.040 & 0.018 \\
 & & 0.50 & 0.050 & 0.048 & 0.022 \\
 & & 0.75 & 0.113 & 0.085 & 0.033 \\
 & & 1.00 & 0.131 & 0.086 & 0.041 \\
\cmidrule{2-6}
 & \multirow{4}{*}{200} & 0.25 & 0.079 & 0.065 & 0.011 \\
 & & 0.50 & 0.165 & 0.094 & 0.026 \\
 & & 0.75 & 0.339 & 0.166 & 0.031 \\
 & & 1.00 & 0.555 & 0.262 & 0.048 \\
\hline
\multirow{8}{*}{LogNormal} & \multirow{4}{*}{50} & 0.25 & 0.033 & 0.043 & 0.002 \\
 & & 0.50 & 0.045 & 0.053 & 0.003 \\
 & & 0.75 & 0.048 & 0.053 & 0.004 \\
 & & 1.00 & 0.050 & 0.054 & 0.004 \\
\cmidrule{2-6}
 & \multirow{4}{*}{200} & 0.25 & 0.044 & 0.044 & 0.009 \\
 & & 0.50 & 0.067 & 0.059 & 0.004 \\
 & & 0.75 & 0.090 & 0.067 & 0.007  \\
 & & 1.00 & 0.138 & 0.082 & 0.011 \\
\hline
\end{tabular}
\end{table}

We validate that the zero-trimmed Mann-Whitney U-test has more power than the other two tests on almost all scenarios of zero-inflated heavy-tailed data, while still controlling type I error.

\subsection{Doubly Robust Generalized U}

\subsubsection{Snapshot DRGU}
\label{apx:DRU_Simu}
We generate $n\in \{50, 200\}$ i.i.d.\ observations $(y_i,z_i,w_i)$ with $p=1$ baseline covariates for simplicity $w_i\sim\mathcal N(0,1)$.  The true propensity score is logistic,
\[
\pi(w_i)=\sigma\!\bigl(-0.2w_i + 0.6w_i^2\bigr),\qquad 
z_i\mid w_i\sim\mathrm{Bernoulli}\!\bigl(\pi(w_i)\bigr),
\]
where $\sigma(x)=1/(1+e^{-x})$. The outcome mean model is: 
\[\mu_0(w_i, z_i)= \beta z_i + 1.0w_i, \qquad y_i\mid(z_i,w_i)\sim\mathcal P\!\bigl(\mu_0(w_i, z_i),\,1\bigr)
\] 
where constant ATE $\beta \in \{0.0, 0.5 \}$ and $\mathcal P$ is one of the normal, log-normal, and Cauchy distributions. We compare Type I error rates and power of correctly specified \texttt{DRGU}, correctly specified linear regression \texttt{OLS}, and Wilcoxon rank sum test \texttt{U} (which does not account for confounding covariates). To probe double robustness, we set up \texttt{misDRGU} as misspecifying the quadratic outcome propensity score model with a linear mean model, while the outcome model in \texttt{misDRGU} is specified correctly. 

\begin{table}[H]
\centering
\caption{Type I Error Rate at sample size = 200}
\label{tab:type1_ugee}
\begin{tabular}{l|cccc}
\hline
Distribution & DRGU & misDRGU & OLS & U\\
\hline
Normal $\alpha=0.05$ & 0.041 & 0.049 & 0.043 & 0.185 \\
\hline
LogNormal $\alpha=0.05$ & 0.054 & 0.070 & 0.054 & 0.150\\
\hline
Cauchy $\alpha=0.05$ & 0.052 & 0.065 & 0.042 & 0.149\\
\hline
Normal $\alpha=0.01$ & 0.014 & 0.005 & 0.012 & 0.045\\
\hline
LogNormal $\alpha=0.01$ & 0.012 & 0.020 & 0.007 & 0.049\\
\hline
Cauchy $\alpha=0.01$ & 0.012 & 0.025 & 0.008 & 0.02\\
\hline
\end{tabular}
\end{table}

\begin{table}[H]
\centering
\caption{Power at $\alpha=0.05$, ATE=0.5}
\label{tab:power_ugee}
\begin{tabular}{lc|cccc}
\hline
Distribution & Sample size & DRGU & misDRGU & OLS & U\\
\hline
\multirow{2}{*}{Normal} & 200 & 0.750 &  0.585 & \textbf{0.940} & 0.299 \\
& 50 & \textbf{0.135} & 0.085 & \textbf{0.135} & 0.035 \\
\hline
\multirow{2}{*}{LogNormal} & 200 & \textbf{0.610} &  0.515 & 0.435 & 0.235 \\
& 50 & \textbf{0.260} & 0.210 & 0.190 & 0.110 \\
\hline
\multirow{2}{*}{Cauchy} & 200 & \textbf{0.660} & 0.580 & 0.435 & 0.310\\
& 50 & \textbf{0.265} & 0.180 & 0.165 & 0.130 \\
\hline
\end{tabular}
\end{table}

\subsubsection{Longitudinal DRGU}
\label{apx:LDRU_Simu}

For the longitudinal setting, we use the same simulation setup as above for observations $(y_{it}, z_i, w_{it})$ for $t=1,...,T=2$ time points. The true propensity score is logistic of time-varying covariates,
\[
\pi(\mathbf w_i)=\sigma\!\bigl(-0.3w_{i1}-0.6w_{i2}\bigr),\qquad 
z_i\mid \mathbf w_i\sim\mathrm{Bernoulli}\!\bigl(\pi(\mathbf w_i)\bigr),
\]
where $\sigma(x)=1/(1+e^{-x})$. The outcome mean model is: 
\[\mu_0(w_{it}, z_i)= \beta z_i + 1.0w_{it}, \qquad y_{it}\mid(z_i,w_{it})\sim\mathcal P\!\bigl(\mu_0(w_{it}, z_i),\,1\bigr)
\] 
We compare three models \texttt{longDRGU}, \texttt{DRGU} using the last timepoint data snapshot, and \texttt{GEE}. The time-varying covariates highlight the strength of using longitudinal method compared to snapshot analysis.

\begin{table}[H]
\centering
\caption{Type I Error Rate at $\alpha=0.05$, sample size = 200, T=2}
\label{tab:type1_lugee}
\begin{tabular}{l|ccc}
\hline
Distribution & longDRGU & DRGU & GEE\\
\hline
Normal & 0.03 & 0.04 & 0.04\\
\hline
LogNormal & 0.04 & 0.05 & 0.02\\
\hline
Cauchy & 0.05 & 0.05 & 0.05\\
\hline
\end{tabular}
\end{table}

\begin{table}[H]
\centering
\caption{Power at $\alpha=0.05$, ATE=0.5, sample size = 200, T=2}
\label{tab:power_lugee}
\begin{tabular}{lc|ccc}
\hline
Distribution & Sample size & longDRGU & DRGU & GEE\\
\hline
\multirow{2}{*}{Normal} & 200 & 0.85 & 0.88 & \textbf{0.92}\\
& 50 & 0.52 & 0.39 & \textbf{0.75}\\
\hline
\multirow{2}{*}{LogNormal} & 200 & \textbf{0.85} & 0.78 & 0.68\\
& 50 & \textbf{0.37} & 0.30 & 0.33\\
\hline
\multirow{2}{*}{Cauchy} & 200 & \textbf{0.83} & 0.76 & 0.66\\
& 50 & \textbf{0.38} & 0.32 & 0.29 \\
\hline
\end{tabular}
\end{table}

\section{Details on A/B Testing}

\subsection{Email Marketing}
\label{apx:Email_AB}
We conducted an A/B test comparing our legacy email marketing recommender system against a newer version designed with improved campaign personalization using neural bandits. We randomly assigned audience members to receive recommendations from either system and measured the downstream impact on conversion value, a proprietary metric measuring the value of conversion.

The resulting conversion value presented challenging statistical properties: extreme zero inflation (>95\% of members had no conversions in both test groups) and significant right-skew among the 1\% who did convert. These characteristics violated the assumptions of conventional testing methods such as the standard $t$-test.

The zero-trimmed Mann-Whitney U-test proved ideal for this scenario by balancing the proportion of zeros between test groups before performing rank comparisons. This approach maintained appropriate Type I error control while providing superior statistical power compared to both the $t$-test and the standard Mann-Whitney U-test. Using the zero-trimmed Mann-Whitney U-test, we detected a statistically significant +0.94\% lift in overall conversion value, most of which was driven by a +0.11\% lift in B2C product conversions among members experiencing the improved campaign personalization (p-value < 0.001). By constast, the $t$-test was able to detect a signficant effect conversion value metric (p-value = 0.249).

\subsection{Targeting in Feed}
\label{apx:LoL_AB}
\begin{figure}[H]
  \centering
  \caption{Distributions of (normalized) impressions and conversions from the targeting in feed experiment.}
  \includegraphics[width=0.45\textwidth]{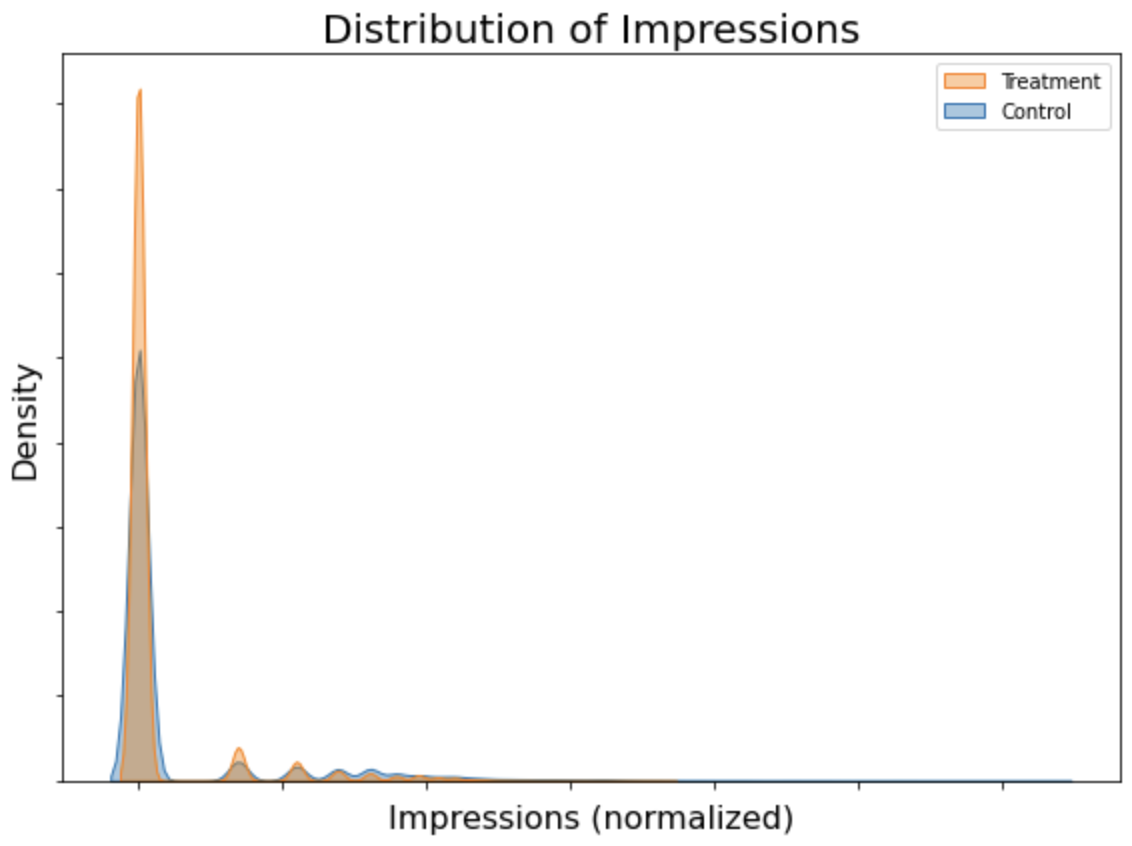}
  \includegraphics[width=0.45\textwidth]{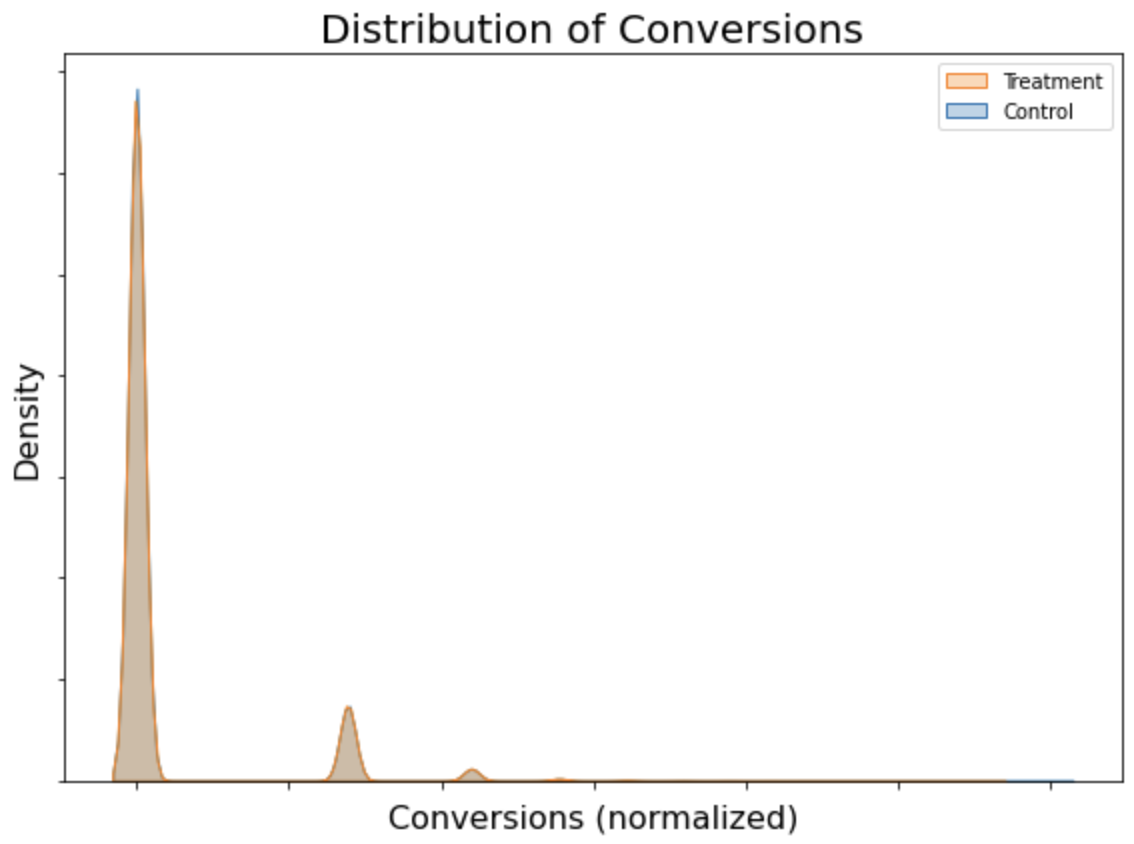}
  \label{fig:lol-imp-conv}
\end{figure}

We conducted an online experiment to evaluate impact of a new marketing algorithm vs legacy algorithm for recommending ads on a particular slot in Feed. The primary interest of the study is downstream conversion impact. Members eligible for a small number of pre-selected campaigns were the unit of randomization. We encountered two main challenges. First, the ad impression allocation mechanism showed a selection bias favoring recommendations from the control system. As a result, we want to adjust for impression as cost and compare return-on-investment (ROI) between the control and treatment group. Second, limited campaign and participant selection introduced potential imbalance in baseline covariates even under randomization. Specifically, we observed that a segment of members with lower baseline conversion rate was more likely to be in the treatment group than in the control group. This introduced the classic case of Simpson's Paradox where conversion rate averaged over all segments is similar in both groups but higher in treatment group when stratified by this confounding segment. We summarized these imbalanced features in Table \ref{tab:lol}. Figure \ref{fig:lol-imp-conv} further shows the large distribution mismatch between impressions in the treatment and control group. We addressed both of these issues by using regression adjustment to estimate lift in ROI while accounting for a confounder such as being in the member segment with low baseline conversion rate. We found the new algorithm to have a statistically significant lift of 1.84\% in conversion per impression, with p-value < 0.001 and 95\% confidence interval (1.64\% - 2.05\%). This is in contrast to failing to reject the null hypothesis of no effect when using two-sample t-test for difference in means of conversion rate (p-value = 0.154).

\begin{table}[H]
\centering
\caption{Characteristics by treatment variant of imbalanced data. Values are relative to mean values in the control group.}
\label{tab:lol}
\begin{tabular}{lcccc}
\hline
 & Control & Treatment\\
 & mean & mean \\
\hline
Conversions & 1.0  & +0.3\% \\
Impressions & 1.0 & -37.7\% \\
Low-baseline segment & 1.0 & +9.5\%\\
\hline
\end{tabular}
\end{table}

\subsection{Paid Search Campaigns}
\label{apx:Paid_AB}
We illustrate leveraging longitudinal repeated measurements in A/B testing (via GEE) to improve power using data collected in an online test run on paid ad campaigns over a 28-day period. We randomized 64 ad campaigns at the campaign level into test and control arms (32 campaigns each), a typical setup for tests run on third-party advertising platforms. We collected daily conversion values for each campaign throughout the experiment, yielding a time series of repeated measurements at the campaign-day level. Due to the limited sample size, a traditional two-sample comparison lacks power to detect the treatment effects in this test. 

To address this small-sample limitation, we fit a Generalized Estimating Equation (GEE) model using campaign as the grouping variable and an exchangeable working-correlation structure to capture within-campaign serial dependence. During the 28-day test, by “borrowing strength” across daily measurements, the GEE framework substantially reduced residual variance and produced tighter confidence intervals around the treatment coefficient. In this phase, the GEE‐estimated treatment effect was very close to significant level (p-value=0.051). In comparison, the snapshot regression analysis using the last snapshot attains p-value at 0.184. 

We also reserved a 28-day validation period prior to the actual launch—during which no treatment was applied—so that treatment and control groups should exhibit no true difference. We collected campaign-day conversion values in the same format and ran the identical GEE analysis, yielding an estimated effect indistinguishable from zero (p-value = 0.82). This confirms that leveraging repeated measurements through GEE both enhances sensitivity to subtle treatment effects and maintains proper control of type I error.

Observing the distribution of the response variables exhibit heavy tail characteristics, we further performed statistical testing using doubly robust U, assuming compound symmetric correlation structure for $R(\alpha)$. We were able to attain statistical significant result with $\hat{P}(y_1>y_0)=0.54$ and $p$-value=0.045.

\end{document}